\documentclass[12pt,letterpaper,oneside, openany]{report}
\usepackage{amsfonts} 
\usepackage{amssymb}  
\usepackage{graphicx} 
\usepackage{amsmath}  
\usepackage{amsmath}  
\usepackage[spanish]{babel}
\usepackage[latin1]{inputenc}
\usepackage{latexsym}
\usepackage{color}
\usepackage{fancyhdr}
\usepackage[T1]{fontenc}
\usepackage{geometry}

\renewcommand{\tablename}{Tabla}

\newcommand{\fig}[1]{Fig~(\ref{#1})}
\newcommand{\tab}[1]{Tabla~\ref{#1}}

\begin{document}
\renewcommand{\tablename}{Tabla}

\pagestyle{empty}
\begin{titlepage}
\begin{center}
{\large \bf UNIVERSIDAD DE LA HABANA.}
\end{center}

\begin{center}
\vspace{0.3cm}

{\large \bf FACULTAD DE FÍSICA.}

\vspace{0.3cm}

\begin{figure}[h!t]
\begin{center}
\includegraphics[width=0.25\textwidth]{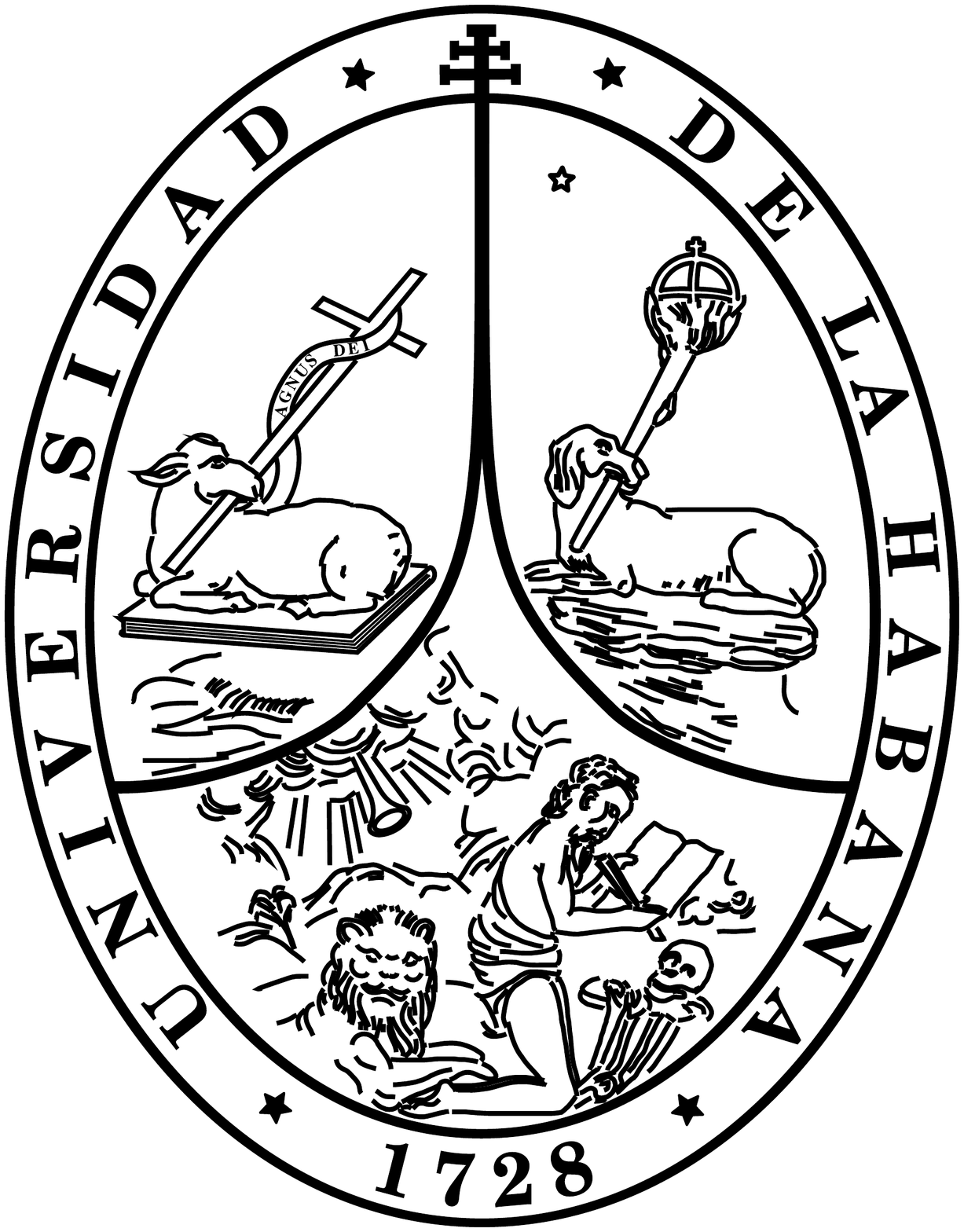}
\end{center}
\end{figure}

\vspace*{0.3cm}

\begin{center}
{\Large \bf Rol del campo magnético en los observables astrofísicos
de Estrellas de Quarks}
\end{center}

\vspace*{1.0cm}

{\large \bf TESIS}\\
{\large  \it presentada en opci\'on al grado de} \\
{\Large  \bf Máster en Ciencias F\'{\i}sicas}\\
\end{center}

\vspace*{1.5cm}

{\large \bf Autor:}\hspace{0.3cm}{\large Lic. Daryel Manreza Paret, Facultad de Física.} \\

{\large \bf Tutor:} \hspace{0.3cm}{\large{Dra. Aurora Pérez
Martínez, ICIMAF.}}

\vspace*{1.5cm}

\begin{center}
\Large{Ciudad de la Habana. Cuba.}\\
\Large{2010}
\end{center}

\end{titlepage}

\newpage
\voffset=1.8in \thispagestyle{empty} \hoffset=3.6in
\noindent {\large{ \emph{...a mi familia}}}\\

%
%

\newpage
\hoffset=0in \voffset=1in \vspace{3in} \thispagestyle{empty} {\large
\bf    Resumen}

Esta tesis tiene el propósito de estudiar cómo se modifican algunos
de los observables astrofísicos de Estrellas de Quarks debido a la
presencia del campo magnético.

Para hacerlo, es necesesario obtener la \textit{Ecuación de Estado}
\textit{(EdE)} y estudiar la \textit{estabilidad} de la Materia
Extraña (formada por quarks $u$, $d$ y $s$) magnetizada densa y fría
en equilibrio estelar (equilibrio $\beta$, conservación del número
bariónico y neutralidad de carga). Trabajaremos  utilizando el
modelo fenomenológico de Bag del MIT.

La \textit{estabilidad} de la Materia Extraña de Quarks Magnetizada
(MEQM) se estudia teniendo en cuenta la variación de los parámetros
más importantes del modelo: la masa del quark $s$, la densidad
bariónica, el campo magnético y el parámetro de Bag. Los resultados
obtenidos se comparan con los de la materia normal de quark
magnetizada (sólo quarks $u$ y $d$ en equilibrio $\beta$) así como
con los de la Materia Extraña de Quarks (MEQ). Se obtiene que la
energía por barión decrece con el aumento del campo magnético lo que
implica que la MEQM es más estable que la MEQ.

Las \textit{EdE} previamente obtenidas se utilizan para obtener
configuraciones estables de  Estrellas Extrañas Magnetizadas,
comprobando que el campo magnético contribuye a disminuir la
relación Masa--Radio (M--R) de la estrella.

\vspace{0.3in}

\hoffset=0in \voffset=1in \vspace{0.5in} \thispagestyle{empty}
{\large \bf Abstract}

This thesis aims to study how amending certain astrophysical
observable of quark stars due to presence of magnetic field.

To do this we need  to obtain Equation of State (EOS) and consider
the stability of Strange Quark Matter (made up of quarks $u$, $d$
and $s$) cold dense and magnetized in stellar equilibrium ($\beta$
equilibrium, conservation of the baryonic number and charge
neutrality). We will work using the phenomenological MIT Bag model.

The stability of the Magnetized Strange Quark Matter (MSQM) is
studies taking into account the variation of parameters from the
model: $s$ quark mass, baryonic density, magnetic field and the Bag
parameter. Results obtained were compared with those of magnetized
normal quark matter (only $u$ and $d$ quarks in $\beta$ equilibrium)
as well as the Strange Quark Matter (SQM). It is found that the
energy per baryon decreases with the increasing magnetic field which
implies that the MSQM is more stable than SQM.

The Equations of State previously obtained are used to obtain stable
configurations of magnetized strange stars checking that the
magnetic field helps to reduce Mass--Radius (M--R) ratio of the
star.

 \vspace{0.3in}



\voffset=0in 

\pagestyle{plain}
\newpage
\voffset=0in 

\pagenumbering{roman} \setcounter{page}{1}

\tableofcontents
\newpage{}
\setcounter{page}{0} \pagebreak \pagenumbering{arabic}
\setcounter{page}{1} \pagestyle{myheadings}

\chapter*{Introducción general.}\label{cap0}
\markright{Introducción general.}




Los Objetos Compactos (OC) son los remanentes que quedan tras la
muerte de una estrella. Dependiendo de sus características se
clasifican en Enanas Blancas (EBs), Estrellas de Neutrones (ENs) y
Huecos Negros (HN)~\cite{MTW}. En estos objetos, la materia está
sujeta a condiciones extremas (altísimas densidades, altas
temperaturas y campos magnéticos muy intensos), ello hace que sean
considerados laboratorios na\-tu\-ra\-les para probar teorías, que
dado el desarrollo tecnológico actual, son imposibles de comprobar
en laboratorios terrestres~\cite{Baym:2006rq}.

En una estrella común existen dos fuerzas que se contraponen de
forma tal que se llega a un estado de equilibrio mecánico. Estas
fuerzas son: la gravitacional, que tiende a comprimir la materia
hacia el centro; y la presión ejercida por la radiación desprendida
en las reacciones de fusión, que tiende a expulsar materia hacia el
exterior de la estrella.

Durante millones de años la estrella quema el combustible nuclear
manteniendo un estado de equilibrio, ubicándose en la llamada
secuencia principal del diagrama de
Hertzsprung--Russell~\cite{Shapiro}. Cuando en la estrella cesan los
procesos de fusión, la gravedad comprime la materia y es la presión
del gas degenerado de electrones, en el caso de las EBs; y de
neutrones, en el de las ENs, lo que garantiza la estabilidad de la
estrella.

La complejidad de Enanas Blancas y ENs  requiere que para
explicarlas dominemos toda la Física que conocemos: la Relatividad
General porque en muy poco espacio está concentrada mucha materia y
el espacio tiempo se deforma de manera apreciable, la interacción
electromagnética porque hay interacción de partículas cargadas que
compensan la carga de la estrella resultando una estrella neutra, la
interacción electrodébil porque hay constante decaimiento de
neutrinos y la fuerte porque o hay quarks deconfinados o hay
hadrones\footnote{Se dividen en bariones(formados por 3 quarks) y
mesones (pares quark--antiquark).} conformados por quarks.

Muchas interrogantes surgen cuando estudiamos  este tipo de objetos.
En par\-ti\-cu\-lar para las ENs  los grandes campos gravitacionales
hacen que las densidades en ellas sean del orden de la densidad del
núcleo atómico y hasta varias veces mayor (la masa del Sol está
concentrada en un radio de
$\sim10\,\mathrm{km}$)~\cite{Lattimer:2004pg}. En la medida que nos
acercamos a su centro, pudieran ocurrir procesos aun más exóticos
que puedan producir que los quarks ``salgan'' de sus neutrones, y se
forme entonces una Estrella de Quarks (EQs).  En realidad sería una
Estrella Extraña porque aparece además de los quarks \emph{up}$(u)$
y \emph{down}$(d)$ otro más masivo: el quarks \emph{strange} o
extraño $(s)$, debido a procesos de desintegración $\beta$ donde se
transforman quarks $u$ y $d$ en $s$~\cite{Bombaci:2008kb}.




Estos objetos serían muy diferentes a las ENs, podrían ser  más
pequeños y compactos, tener  unos $6 \mathrm{km}$ de radio y rotar
más rápido. La materia estaría cohesionada debido a la interacción
fuerte y no a la gravitatoria como ocurre en las
ENs~\cite{SchaffnerBielich:2004ch}.

Por tanto, conocer qué compone el interior de una ENs o afirmar que
la existencia de una EQs es una realidad, está condicionado a
responder cuál es el estado fundamental de la materia a muy altas
densidades y con ello responder a la pregunta, desde la física de
partículas elementales: ¿qué le ocurre a la materia cuando es
``comprimida''  hasta alcanzar densidades por encima  de la densidad
nuclear
($n_0=0.16\,\mathrm{fm}^{-3}$)\,\,\footnote{$1\,\mathrm{fm}=10^{-13}\,\mathrm{cm}$}?


Hoy en día, la teoría aceptada para describir el comportamiento de
las partículas subatómicas y sus interacciones es el Modelo
Estándar. En este modelo se utiliza el aparato matemático de la
Teoría Cuántica de Campos para unificar el Modelo Electrodébil
(interacción electromagnética y débil unificada) y la Cromodinámica
Cuántica (CDC) en una teoría que describe las interacciones de todas
las partículas observadas experimentalmente.

Según el Modelo Estándar, la materia está compuesta por doce
fermiones: seis quarks y seis leptones\footnote{neutrinos y
electrones} (y sus antipartículas), además están los bosones
mediadores de las interacciones. En la \tab{partículas} se muestran
algunas de las propiedades de los quarks~\cite{Roy:1999rc}.
\begin{table}[h!]
\begin{center}
\begin{tabular}{|l|c|c|}
\hline

Quarks&Masa $(\mathrm{MeV})$&Carga($e^-=-1$)\\
\hline\hline

u~(up)&$1.5-4.0$& $+2/3$\\ \hline

d~(down)&$4-8$& $-1/3$\\ \hline

s~(strange)&$80-130$&$-1/3$ \\ \hline

c~(charm)&$1150-1350$& $+2/3$\\ \hline

t~(bottom)&$4100-4400$& $-1/3$\\ \hline

b~(top)&$170900\pm1800$& $+2/3$\\

\hline
\end{tabular}
\end{center}
\caption{\small Masas y cargas de los quarks~\cite{PDG}.}
\label{partículas}
\end{table}

Los quarks son los ``ladrillos'' que junto con los electrones forma
la materia tal y como la conocemos: los nucleones (neutrones y
protones), que están formados por quarks $u$ y $d$ y junto a los
electrones forman átomos y estos las moléculas. El hecho de que
ellos estén unidos dentro de los nucleones se debe a la interacción
fuerte  o de color (CDC) que opera a cortas distancia $\sim
10^{-13}\,\mathrm{cm}$.


La CDC  describe las interacciones entre los quarks y los gluones,
partículas mediadoras de la interacción fuerte, análogas a los
fotones (mediadores de la interacción electromagnética). Entre las
principales características que se derivan de la CDC están el
confinamiento y la libertad asintótica que dan  cuenta de que no
hayan quarks libres en la Naturaleza, estén confinados a los
hadrones y  se comporten como libres dentro de ellos.

\begin{figure}[t]
      \centering
      \includegraphics[height=8cm,width=9cm]{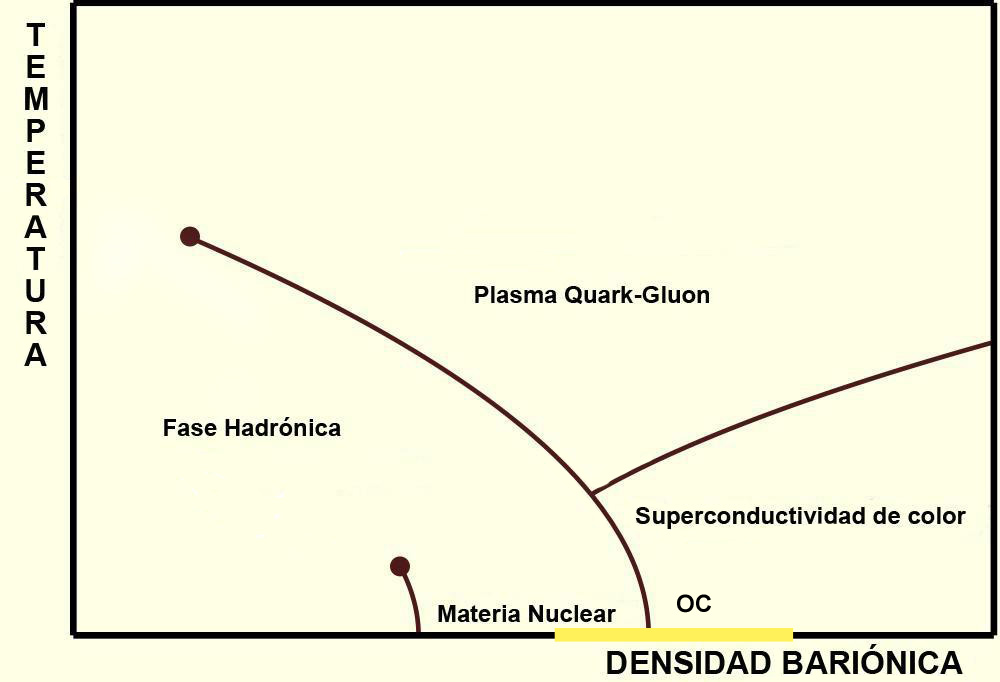}
      \caption{Esquema del diagrama de fase de la CDC}
\label{QCDphase}
\end{figure}

Sin embargo la CDC concibe que en condiciones extremas: alta
temperatura o alta densidad, pueden existir quarks
libres~\cite{Bodmer:1971we,Witten:1984rs}. Por tanto, pensar que el
interior de una ENs  pudiera ser el escenario real de la existencia
de este tipo de materia es totalmente plausible. En el diagrama de
fase de la CDC \fig{QCDphase} se representa la zona en que
``habitan'' los OC, es decir, bajas temperaturas ($T\sim
30\,\mathrm{MeV}$) y altos potenciales químicos ($\mu
>300\,\mathrm{MeV}$).

Los radios y masas de las EQs o las ENs se pueden medir
indirectamente. Ellas emiten radiación Gamma y rayos X que se
detectan por instrumentos colocados en
satélites~\cite{Lattimer:2006xb}. A estas estrellas se les asocian
además campos magnéticos muy intensos que se explican por la
amplificación del campo de las estrellas progenitoras después de la
explosión de  la Supernova. Estos campos son extraordinariamente
grandes: de $10^{15}$ Gauss (el campo terrestre es de 0.5 Gauss) en
la superficie, y $10^{18}$ Gauss $(\mathrm{G})$ en el
interior~\cite{Shapiro}. La NASA en el 2002 reportó 3 fuentes
candidatas a Estrellas Extrañas \cite{weber}.


En esta  tesis nos proponemos estudiar la Materia de Quarks (MQ) en
presencia de campos magnéticos intensos en equilibrio estelar y los
observables astrofísicos que se obtienen para EQs formadas por MQ
magnetizada.

Este trabajo le da entonces continuidad a
\cite{Felipe:2007vb,Felipe:2008cm,PerezMartinez:2005av} donde fueron
estudiadas las propiedades termodinámicas del gas de quarks
magnetizado porque, no solo se estudia la estabilidad de la materia
de quarks y las EdE en condiciones realistas, sino también se
investigan las implicaciones macroscópicas que tendría la existencia
de EQs formadas por este tipo de materia magnetizada. Esto último se
hace tomando en cuenta los efectos de la Relatividad General en el
equilibrio hidrodinámico de la estrella a través de las ecuaciones
Tolman-Oppenheimer-Volkoff (TOV) ~\cite{Oppenheimer}.

Para ello usaremos el Modelo Fenomenológico MIT
\footnote{Massachusetts Institute of Technology.} Bag Model que
permite describir el confinamiento de los quarks sin necesidad de
usar la Cromodinámica Cuántica $CDC$ que introduce grandes
complicaciones matemáticas. En dicho modelo los quarks son
considerados libres dentro de un volumen (Bag) y es precisamente
este quien reproduce el confinamiento. El equilibrio estelar
significa que consideraremos que transcurre suficiente tiempo como
para que ocurran procesos de desintegración $\beta$ dentro de la
estrella, mientras que la misma se mantiene eléctricamente neutra y
conserva el número de bariones en su interior.

Un estudio así tiene  importancia no solo para la Astrofísica sino
también para la Física de Partículas. Buscando concordancia con las
observaciones podríamos restringir parámetros de los modelos
teóricos que influirían en las propiedades de la física del
micromundo, hasta que lleguen experimentos terrestres que serían los
más conclusivos.

En la tesis nos proponemos concretamente los siguientes objetivos:
\begin{enumerate}
\item Obtener la EdE, relación entre la
presión y la densidad de energía de la MEQ y la materia normal de
quarks (materia eléctricamente neutra formada por quarks $u$ y $d$
en equilibrio $\beta$ con electrones), en presencia de campos
magnéticos intensos en equilibrio estelar.
\item Estudiar la estabilidad de la MEQM en equilibrio estelar.
Probar que la energía por barión para la MEQM es menor que para la
MEQ y para la materia normal  de quarks magnetizada.
\item Resolver las ecuaciones de TOV usando las EdE previamente obtenidas y así
determinar las configuraciones estables de masa-radio para las
estrellas de quark extrañas magnetizadas. Obtener además la masa
bariónica y el corrimiento al rojo gravitacional para EQs.
\end{enumerate}

%

%



La tesis está organizada de la siguiente forma: Los
\textbf{Capítulos~\ref{cap1} y~\ref{cap2}} están dedicados a
introducir los aspectos fundamentales referente a los objetos
compactos y a las propiedades de la materia de quarks
respectivamente. En ellos sólo pretendemos dar a conocer los
principales avances en estas áreas, por lo cual profundizaremos solo
en los aspectos que nos sean necesarios en los demás capítulos. En
el \textbf{Capítulo~\ref{cap3}} exponemos las principales
características de la materia de quarks magnetizada, obtenemos las
EdE y estudiamos la estabilidad de la materia de quarks magnetizada
en equilibrio estelar. En el \textbf{Capítulo~\ref{cap4}} obtenemos
las configuraciones estables de masa y radio para EQs, así como la
masa bariónica y el corrimiento al rojo gravitacional.
Posteriormente se exponen las conclusiones y recomendaciones de la
tesis.




\chapter{Objetos Compactos: Estrellas de Neutrones, Estrellas Extrañas.}\label{cap1}
\markright{Capítulo 1: Objetos Compactos: Estrellas de Neutrones,
Estrellas Extrañas.}

\section{Características generales.}
\emph{En el Universo existe una gran variedad de fenómenos que
despiertan el interés y el asombro de los seres humanos, sin dudas
las estrellas constituyen uno de los escenarios más atractivos  para
la investigación.}

\begin{figure}[t]\label{masaENs}
     \centering
     \includegraphics[height=4.2in,width=3.0in]{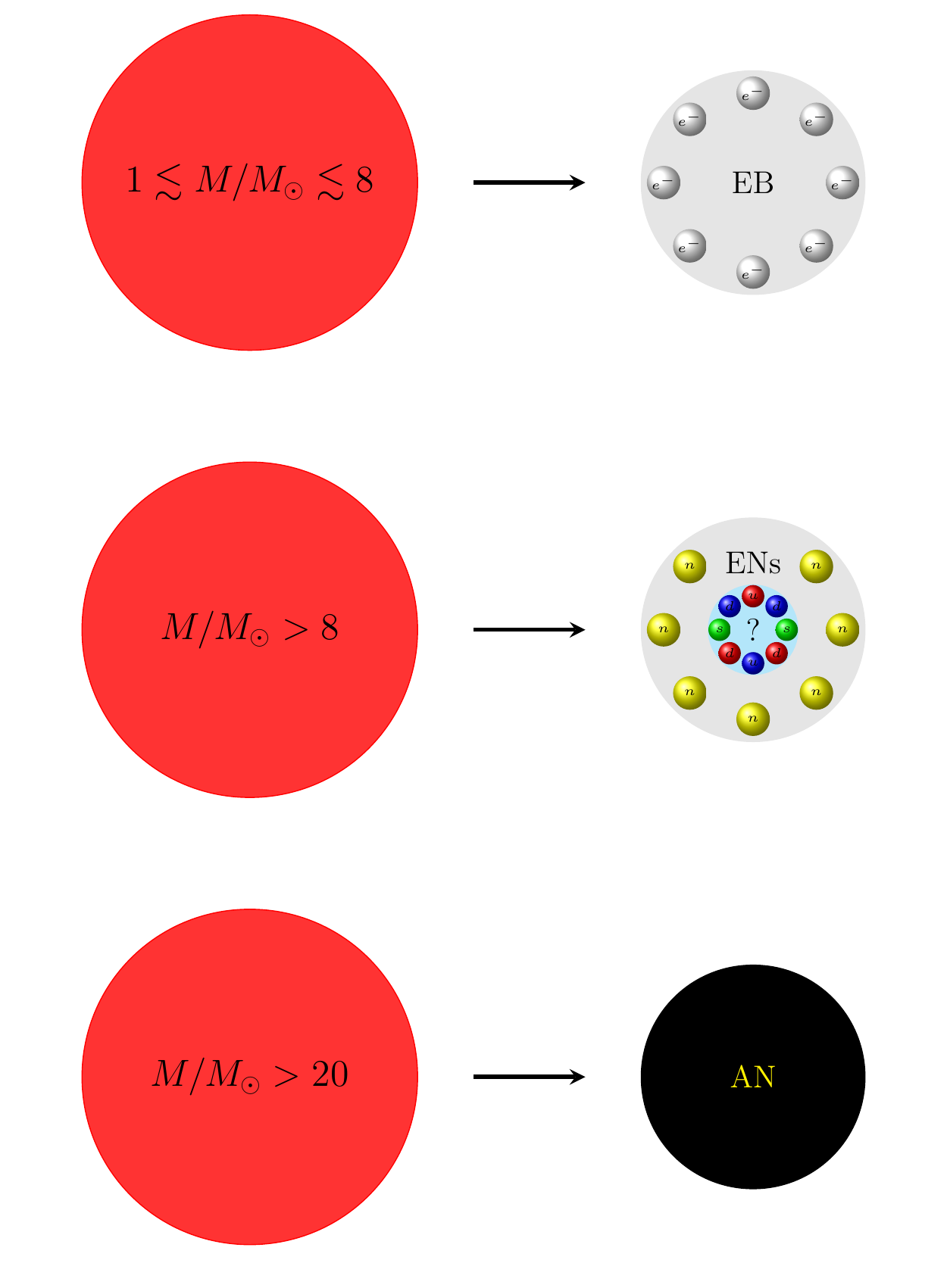}
     \caption{\small Diagrama que muestra el destino final de las
     estrellas en dependencia de su masa~\cite{Ruester:2006yh}.}
\end{figure}

La vida de una estrella pasa por diferentes etapas. Todo comienza
cuando las nubes de gas estelar comienzan a agruparse
gravitacionalmente, la temperatura aumenta y comienzan a fusionarse
los elementos ligeros como el hidrógeno. En estas reacciones de
fusión se van formando elementos más pesados. Cuando el combustible
nuclear se agota, la estrella se transforma en una Supernova, su
destino final depende en gran medida de la masa inicial de la
estrella~\cite{Ruester:2006yh}. En la \fig{masaENs} se resumen los
posibles destinos de una estrella en dependencia de su masa, y sus
principales características se resumen en la \tab{propOC}.

Entre los posibles destinos finales de las estrellas, nos
centraremos en las ENs, debido a que como ya mencionamos con
anterioridad ellas podrían tener en su interior una fase formada por
MEQ formando o una Estrella Extraña o una híbrida.
\begin{table}[h!]
\begin{center}
\begin{tabular}{|l|l|l|c|}
\hline

Objeto Compacto&Masa$^a$ $(M)$&Radio$^b$ $(R)$&Densidad Media $(\,\mathrm{g}\,\,\mathrm{cm}^{-3})$ \\
\hline\hline

Enana Blanca& $\lesssim1.4M_{\odot}$&$\sim10^{-2}R_{\odot}$& $\lesssim10^7$  \\
\hline

Estrella de Neutrones& $\sim 1-3M_{\odot}$&$\sim10^{-5}R_{\odot}$&$\lesssim10^{15}$\\
\hline

Hueco Negro& Arbitraria& $2GM/c^2$& $\sim M/R^3$ \\

\hline

\multicolumn{2}{l}{\small$^a \text{Masa del Sol}\, M_{\odot}=1.989\times10^{33}\,\mathrm{g}$} \\
\multicolumn{2}{l}{\small$^b \text{Radio del Sol}\,
R_{\odot}=6.9599\times10^{10}\,\mathrm{cm}$}

\end{tabular}
\end{center}
\caption{\small Características de los Objetos
Compactos~\cite{Shapiro}.} \label{propOC}
\end{table}

\section{Estrellas de Neutrones.}
Las ENs fueron predichas por Landau poco tiempo después de haberse
descubierto el neutrón en el año 1932, desde entonces mucho se ha
avanzado en el estudio de sus propiedades. Una ENs es el remanente
de una estrella supergigante roja que al agotarse los elementos
necesarios para los procesos de fusión comienza a colapsar,
eyectando las capas externas en una explosión conocida como una
Supernova tipo II. La materia que queda en el centro de la estrella
se encuentra sometida a campos gravitacionales muy elevados. Debido
a las altas densidades, los electrones son capturados por los
protones formándose neutrones en una reacción $\beta$ inversa
$(p+e^-\rightarrow n+\nu_e)$. La presión del gas degenerado de
neutrones es la que frena el colapso de la estrella equilibrando de
esta manera la fuerza gravitacional~\cite{Lattimer:2004pg}.

Las ENs están formadas por capas en las cuales podemos encontrar
distintas familias de partículas. Se pueden identificar cinco capas
fundamentales~\cite{Lattimer:2004pg}: el núcleo interior y exterior,
la corteza, la envoltura y la atmósfera. En la atmósfera y la
envoltura se encuentra una cantidad despreciable de masa de la
estrella, estas capas están formadas principalmente por electrones,
núcleos y átomos. La corteza se extiende por unos $2\,\mathrm{km}$,
está compuesta por núcleos que varían de acuerdo a la densidad, van
desde el núcleo de $^{56}\mathrm{Fe}$ para densidades de $10^6\,
\mathrm{g}\, \mathrm{cm}^{-3}$ hasta núcleos con número bariónico
$A\sim200$ para densidades $n\approx n_0/3$. A medida que la
densidad continúa aumentando en la corteza hay toda una serie de
cambios de fase hasta llegar al núcleo donde se concentra el $99\%$
de la masa. El núcleo exterior es una mezcla de nucleones,
electrones y muones. En el núcleo interior se espera que exista una
transición de fase a materia deconfinada de quarks.

Una característica importante de las ENs son sus elevados campos
magnéticos $(B\sim10^{12}-10^{14}\,\mathrm{G})$, el mecanismo
mediante el cual surgen estos campos no se comprende del todo. El
modo más simple de explicar su surgimiento es producto de la
compresión del flujo magnético durante el colapso del núcleo de la
estrella progenitora, con esto se logra explicar la formación de
campos entre $10^{11}-10^{13}\,\mathrm{G}$.

Hay otro efecto que ocurre dentro de algunas ENs que pudiera
explicar campos magnéticos mil veces mayores. Dentro de ellas el
``gas de partículas''circula por convección, es decir, la
temperatura no es uniforme luego hay partes del gas que están más
calientes que otras. Como el gas es un buen conductor eléctrico
cualquier línea de campo magnético lo arrastra moviéndolo con él. De
esta manera el campo puede amplificarse. Este efecto se conoce como
el Efecto Dinamo~\cite{Duncan1,Duncan2} y se piensa que es el que
origina el campo magnético en todas las estrellas e incluso en
planetas; está presente durante toda la vida de las estrellas de una
manera ``sedada'' pero para las EN cuyo núcleo rota rápidamente la
convección llega a ser muy violenta. La primera simulación numérica
hecha para ENs a partir de un modelo con convección puede verse
en~\cite{Lattimer} y los resultados muestran que si una Estrella
Neutrónica naciente tiene una temperatura de $3 ×
10^{9}\,\mathrm{K}$ el fluido caliente circula en diez milisegundos
y luego cesa. Esta furiosa convección explica el magnetismo de este
tipo de estrella de neutrones~\cite{Duncan1}. En otras palabras,
este modelo nos explicaría que si una estrella naciente de neutrones
entrega el 10\% de su energía cinética al campo magnético este puede
llegar a $10^{15}$ Gauss que es $10^{3}$ veces superior a los campos
que se deducen que tengan los radio pulsares. Entonces, si el efecto
dinamo opera globalmente dependerá de la relación
rotación-convección. Así una ENs con períodos de rotación del orden
o superiores al período de convección de 10 milisegundos logrará
campos de $10^{15}$ Gauss. Ellas son las llamadas Magnetars
(Magnetic Stars-Estrellas Magnéticas). Por otro lado el Efecto
Dinamo no tendrá lugar en pulsares como el Pulsar del Cangrejo que
rota una vez cada 20 milisegundos, mucho menor que el período de
convección. En ese caso el campo magnético no se incrementaría mas
allá de $10^{12}$ Gauss.

Hemos visto dos mecanismos que de conjunto podrían explicar campos
magnéticos intensos en las ENs llamadas Magnetars. No obstante a
pesar de que estas teorías son reconocidas por la comunidad
científica no se descartan otras ideas para explicar los colosales
campos magnéticos que en ENs aparecen~\cite{12}.

\section{Estrellas de Quarks.}

El primer modelo de EQs fue propuesto por Itoh en el año
1970~\cite{itoh}, con este modelo simple se investigó el equilibrio
hidrostático de las EQs siguiendo el método de Oppenheirmer y
Volkoff. La MEQ puede encontrarse como una fase en el núcleo de las
ENs en cuyo caso tendremos una estrella híbrida, o puede formar una
EQs, es decir, una estrella formada en su totalidad por quarks con
una pequeña capa exterior de electrones.

Debido a la presencia de la fuerza nuclear fuerte las EQs son
sistemas autoligados, a diferencia de la materia nuclear, que está
ligada mediante la gravedad. Esto hace que las EQs puedan presentar
configuraciones más compactas, es decir, menores radios con igual
masa.

Las EQs al ser autoligadas no presentan una masa mínima como las ENs
$(M_{min}=0.1\,M_{\odot})$~\cite{Shapiro}, y su masa máxima tiene
una dependencia con el radio máximo (para masas pequeñas en
comparación con la masa máxima) del tipo $M\sim R^3$, en contraste
con las ENs, cuyo radio decrece con el aumento de la masa $(M\sim
R^{-3})$.

Otra característica que distingue a las EQs es que violarían el
límite de Eddington~\cite{Lattimer:2006xb}. El límite de Eddington
implica la existencia de una luminosidad máxima para una estrella
con una masa dada, tal que la configuración se encuentre en
equilibrio hidrostático. Para una estrella normal, el límite de
Eddington está dado por
$$L_{\rm Edd}\simeq 8.1 \times 10^{43}\frac{M}{M_\odot} \,\mathrm{MeV\,s}^{-1}\,.$$

Para cualquier valor de radiación que supere este límite, no habrá
equilibrio hidrostático, causando la pérdida de masa de la estrella.
El mecanismo de emisión en una EQs produciría luminosidades por
encima de dicho límite. Al ser autoligada la EQs, el límite de
Eddington no le es aplicable. Así, su superficie puede alcanzar
temperaturas mucho mayores que las que se observan en una ENs, y por
lo tanto, la emisi\'on de energía t\'ermica~\cite{Usov} también debe
ser ma\-yor.

\emph{A pesar de estas diferencias  experimentalmente es difícil
diferenciar a una ENs de una EQs.}

\section{Evidencias observacionales.}

La primera ENs fue descubierta en 1967 por Jocelyn Bell al estudiar
las señales de radio provenientes de un objeto en el cielo. Jocelyn
descubrió el pulsar CP 1919 (Cambridge Pulsar $19^h19^m$) observando
que la señal de radio que recibía tenía un período exacto de 1.3373
segundos y un ancho de pulso de 0.04 segundos~\cite{Hewish}. Al
inicio se pensó incluso que esta señal provenía de alguna
civilización extraterrestre, hasta que Thomas Gold y Fred Hoyle
identificaron que la señal se generaba a partir de una ENs en
rotación con un elevado campo magnético~\cite{Gold}.

En la actualidad han sido identificados más de 2000 objetos como
ENs. La determinación de parámetros como la masa, el radio y las
características de enfriamiento pueden imponer restricciones a las
EdE para la materia a altas densidades. Los astrónomos buscan
fundamentalmente estrellas aisladas para determinar sus propiedades,
a partir de los espectros de emisión, sin que interfieran
complicados mecanismos de transmisión de materia de un objeto a
otro~\cite{Drake:2002bj}.

La mejor forma de determinar la masa de una ENs es a través de la
tercera ley de Kepler~\cite{Shapiro}. Uno de los mejores resultados
en la determinación de masas se obtuvo para el pulsar binario PSR
1913+16 cuyas masas son $1.3867\pm0.0002$ y
$1.4414\pm0.0002\,M_{\odot}$~\cite{Weisberg:2004hi} respectivamente.

La determinación de los radios se hace de forma indirecta, se
combinan los datos observacionales con la teoría para obtener
algunos estimados que arrojan valores entre
$9-15\,\mathrm{km}$~\cite{Lattimer:2006xb}.

El campo magnético se determina a partir de las observaciones del
decrecimiento en el período de rotación de los pulsares.  Aunque
estas mediciones no son muy precisas, se estiman valores de entre
$10^{12}$ y $10^{15} \,\mathrm{G}$~\cite{Xu:2001bp}.

El corrimiento al rojo $z_s$ se determina a partir del análisis del
espectro de emisión de la estrella.

Existe toda una serie de observaciones que hacen pensar que algunos
de los objetos que se comportan como ENs, son en realidad EQs o
estrellas híbridas.

El observatorio Chandra de rayos X de la NASA encontró dos estrellas
inusuales: la fuente RX J1856.5-3754~\cite{Xu:2002ns,Trumper:2003we}
con una temperatura de $10^5\, \mathrm{K}$ y la fuente 3C58  con un
período de 65 ms. RX J1856.5-3754 es demasiado pequeña para ser una
ENs convencional y 3C58 parece haberse enfriado demasiado rápido en
el tiempo de vida que se le estima.

Combinando los datos del Chandra y del telescopio espacial Hubble,
los astrónomos determinaron que RX J1856.5-3754 radia como si fuera
un cuerpo sólido con una temperatura de unos $7\times 10^5
\,\,^o\mathrm{C}$ y que tiene un diámetro de alrededor de 11 km, que
es un tamaño demasiado pequeño como para conciliarlo con los modelos
conocidos de las ENs.

La fuente de rayos X denominada EXO 0748-676 es otro de los objetos
que atrae la atención pues posee una masa de $1.5<M/M_{\odot}<2.3$ y
un radio de $9.5\,\mathrm{km}<R<15\,\mathrm{km}$ con un corrimiento
al rojo de $z_s=0.35$~\cite{Ozel:2006km}.

En febrero de 1987 en el observatorio de Las Campanas, en Chile, fue
observada la Supernova 1987A en la Gran Nube de Magallanes.  El
remanente estelar que ha quedado como consecuencia de la explosión
de esta Supernova, podría ser una EQs, ya que el período de emisión
de este pulsar es de $P_{em} = 0,5 \mathrm{ms}$. Una ENs canónica no
podría tener una frecuencia de rotación tan alta
~\cite{weber,Xu:2007wd}.








\chapter{Materia de quarks.}\label{cap2}
\markright{Capítulo 2: Materia de quarks.}

\noindent

\section{Modelo para describir la materia de quarks.}

\subsection{CDC y materia de quarks.}

El Lagrangiano de la CDC es~\cite{Buballa}:
\begin{equation}\label{CDC_Lagr}
   {\cal L}_{QCD} \;=\; \bar q \,(\,i\gamma^\mu D_\mu - \hat m\,)\,q
    \;-\; \frac{1}{4}\, G^{a\,\mu\nu}\, G^a_{\;\mu\nu}~,
\end{equation}
donde $q$ es el campo de los quarks con seis sabores ($u,d,s,c,b,t$)
y tres colores, y  $\hat m = {\rm diag}_f(m_u, m_d, \dots)$ es la
matriz de masa. La derivada covariante
\begin{equation}
D_\mu \;=\; \partial_\mu \;-\;i g \,\frac{\lambda^a}{2}\,A_\mu^a
\end{equation}
está relacionada con el campo de los gluones $A_\mu^a$ y
\begin{equation}
G^a_{\;\mu\nu} \;=\; \partial_\mu\,A^a_\nu
\;-\;\partial_\nu\,A^a_\mu
    \;+\; g\,f^{abc}\,A_\mu^b\,A_\nu^c
\label{Gamn}
\end{equation}
es el tensor de fuerzas de los gluones. $\lambda^a$ y $f^{abc}$ son
los generadores del grupo $SU(3)$ (matrices de Gell-Mann) y las
correspondientes constantes antisimétricas de estructura
respectivamente. $g$ es la constante de acoplamiento de la CDC.

El Lagrangiano~(\ref{CDC_Lagr}) es por construcción simétrico bajo
las transformaciones del grupo $SU(3)$ en el espacio de color. Este
grupo tiene un carácter no Abeliano debido a la presencia de los
términos $f^{abc}$ en~(\ref{Gamn}). Debido a esto, como ya se
explicó en la introducción, la CDC presenta algunas características
peculiares, que no están presentes en teorías de calibración
Abelianas como la Electrodinámica Cuántica, ellas son:
\begin{itemize}
  \item Los gluones presentan carga de color.
  \item La CDC es una teoría asintóticamente libre, es decir el
  acoplamiento disminuye para cortas distancias o de forma
  equivalente para grandes energías. En la aproximación de un lazo
  tenemos que:
  \begin{equation}\label{alpha}
    \alpha_s\equiv\frac{g_s^2}{4\pi}\approx\frac{4\pi}{(11-\frac{4}{3}N_f)ln(\mu^2/\Lambda^2_{CDC})},
  \end{equation}
  donde $N_f$ es el número de sabores y $\Lambda_{CDC}=(200\sim300) \,
  MeV$ es un parámetro de renormalización.
  \item Para bajas energías la interacción es fuerte,
  es decir, la fuerza entre las partículas aumenta con la distancia (confinamiento).
  \item El Lagrangiano presenta (aproximadamente) simetría quiral,
  es  decir, es simétrico ante las transformaciones globales $SU(N_f)_L \times
  SU(N_f)_R$. Esta simetría sería exacta en el límite en que el número de sabores de quarks
  $N_f$,  tenga masa nula.
\end{itemize}

Veamos con más detenimiento las implicaciones de la
ecuación~(\ref{alpha})~\cite{Xu:2008nd}.  En el caso de la
Electrodinámica la constante de acoplamiento (constante de
estructura fina) tiene un valor $\alpha_{EM}\simeq1/137<0.01$, este
hecho hace que podamos dar un tratamiento perturbativo a esta
teoría.

Para  la CDC se podrían utilizar métodos perturbativos en el límite
$\mu\rightarrow\infty$, esto equivale, para la  MEQ a la condición
de que la densidad bariónica tienda a infinito
$(n_B\rightarrow\infty)$ ya que el potencial químico bariónico sería
$\mu\simeq(3\pi^2)^{1/3}\hbar cn_B^{1/3}$.

Estimemos el valor de $\alpha_s$ para las densidades que se
encuentran en el interior de las EQs. Para  densidades $n_B=10n_0$,
obtenemos que $\alpha_s\simeq0.8$. Incluso si $n_B=10^6n_0$
obtenemos que $\alpha_s\simeq0.15$. Para que
$\alpha_s\sim\alpha_{EM}$ se necesitaría que $n_B>10^{123}n_0$.

Estas estimaciones nos aseguran la imposibilidad de utilizar Teoría
de Perturbaciones a partir de la CDC si intentamos describir un
sistema real, como es el interior de los OC. Por ello para estudiar
estos sistemas se han propuesto varios modelos fenomenológicos, uno
de los más exitosos es el Modelo de Bag del MIT, que describiremos
seguidamente y que será el que usaremos en esta tesis.

\subsection{Modelo de Bag del MIT.}

El modelo de Bag (bolsa) del MIT  fue propuesto en los años
70~\cite{Chodos} para describir desde el punto de vista microscópico
a los hadrones. En este modelo se logra reproducir la libertad
asintótica y el confinamiento de los quarks, dos de las propiedades
más importantes de la CDC, a través de un parámetro libre del modelo
$B_{bag}$.

En este modelo se describe a los quarks como partículas libres (o
casi libres) confinados dentro de un Bag. La estabilidad del Bag se
garantiza introduciendo el parámetro $B_{bag}$ que se interpreta
como una contribución positiva a la densidad de energía, y negativa
a la presión dentro del volumen. De forma equivalente podemos
atribuir el término $-B_{bag}$ a la región fuera del Bag, lo que
conduce a un vacío con una densidad de energía negativa
$\epsilon_{vac}=-B_{bag}$ y una presión positiva $p_{vac}=+B_{bag}$.
El tensor energía impulso dentro del Bag para un fluido perfecto
viene dado por:
\begin{equation}\label{Tensor EM}
\mathcal{T}^{\mu}_{\,\,\,\,\nu}=diag(\epsilon,P,P,P),
\end{equation}
donde $\epsilon$ es la densidad de energía y $P$ es la presión del
fluido.

El término $B_{bag}\,g^{\mu}_{\,\,\,\,\nu}$ (siendo
$g^{\mu}_{\,\,\,\,\nu}$ el tensor métrico) se añade al tensor
energía impulso dentro del Bag quedando entonces las siguientes
expresiones para la presión y para la densidad de energía:
\begin{equation}\label{pres&energia_bag1}
     P = \sum_{f}P_{f}-B_{bag}
\end{equation}
\begin{equation}\label{pres&energia_bag2}
  \epsilon=\sum_{f}\epsilon_{f}+B_{bag},
\end{equation}
donde la suma es sobre los sabores de los quarks.


El modelo fenomenológico de Bag del MIT es ampliamente usado para
describir la materia en el interior de los OC. Es el que
utilizaremos en el Capítulo~\ref{cap3} para encontrar las EdE de la
MEQM y estudiar la estabilidad de la misma.

\section{Hipótesis de la materia extraña de Quarks.}

La hipótesis sobre la existencia de la MEQ o conjetura de
Bodmer--Witten establece que la materia formada por quarks $u$, $d$
y $s$ en estado libre, puede ser el estado básico de la materia.
Este hecho se cumple si se satisface la desigualdad:
\begin{equation}\label{des_uds1}
   \left.\frac{E}{A}\right|_{MEQ}<\left.\frac{E}{A}\right|_{^{56}\text{Fe}},
\end{equation}
es decir, la energía por barión para la MEQ es menor que la energía
por barión para el $^{56}\text{Fe}$ (núcleo más estable de la
naturaleza).

El hecho empírico de que en la Naturaleza la materia que se observa
está formada por núcleos compuestos por neutrones y protones, y que
no hallemos la MEQ, no entra en contradicción con la hipótesis de su
existencia. Para que ésta se forme a partir de núcleos normales es
necesario que los quarks $u$ y $d$, que forman los neutrones y
protones, se transformen en quarks $s$. Pero no solo con eso basta,
se necesita para lograr la estabilidad que la fracción de quarks $s$
sea grande $(n_s\approx n_u\approx n_d)$. La probabilidad de que
ocurra la reacción débil que garantiza que se formen los quarks $s$
es casi nula, lo que hace que solo para grandes densidades, en el
interior de las ENs y durante millones de años, se forme la cantidad
necesaria de quarks $s$ para que la MEQ sea más estable que la
materia nuclear \cite{Madsen:1999ci}.


Sin embargo, la estabilidad de los núcleos sí excluye la estabilidad
de la materia compuesta solamente por quarks $u$ y $d$, como se
demuestra a continuación.

\subsection{Estabilidad de la materia de quarks normal en equilibrio
estelar.}\label{estmatud}

Analicemos la estabilidad para el caso de la materia de quarks
formada por quarks $u$ y $d$ en equilibrio $\beta$ con electrones.
Este tipo de materia pudiera formarse en el interior de los OC al
formarse el plasma quark--gluón. Como veremos, la misma constituiría
un estado metaestable, pues tiene mayor energía por barión que la
materia nuclear.

El equilibrio químico dentro de la estrella se garantiza a través de
la reacción de desintegración $\beta$:
\begin{equation}\label{eq_beta_ud}
d \leftrightarrow u+e^-+\bar{\nu_e}
\end{equation}
Como las estrellas son neutras se impone la condición de neutralidad
de carga que en el caso general se expresa mediante:
\begin{equation}\label{Nutralidad}
   \sum_{f}n_fq_f-n_e=0,
\end{equation}
donde $q_f\,\,(f=u,d,s)$ son las cargas de los quarks que vienen
dadas en la \tab{partículas} y $n_i\,\,(i=u,d,s,e)$ (emplearemos el
subíndice $f$ cuando tengamos en cuenta solamente los quarks y el
subíndice $i$ cuando incluyamos a los electrones) son las densidades
de las partículas involucradas. Se impone también la conservación
del número bariónico:
\begin{equation}\label{cons_nB}
    n_B=\frac{\sum_{f}n_f}{3}
\end{equation}

De las condiciones anteriores podemos escribir el siguiente sistema
ecuaciones:
\begin{subequations}\label{sist_ud}
\begin{eqnarray}
\mu_u+\mu_e-\mu_d&=&0  \ \ \  \text{equilibrio}\,\beta \\
2n_u-n_d-3n_e&=&0    \ \ \   \text{neutralidad de carga}\\
n_u+n_d-3n_B &=&0     \ \ \   \text{conservación del número
bariónico},
\end{eqnarray}
\end{subequations}
donde $\mu_i$ y $n_i$ son los potenciales químicos de las
partículas.

Hemos supuesto que ha pasado el tiempo suficiente para que los
neutrinos hayan escapado del sistema por lo que tomaremos
$\mu_{\bar{\nu}_e}=0$. El sistema~(\ref{sist_ud}) es un sistema de
tres ecuaciones con tres incógnitas $\mu_i\,\,(i=u,d,e)$. Los
resultados obtenidos a partir de este sistema serán utilizados para
comparar con los de la MEQ.


Podemos realizar un análisis de la estabilidad de este tipo de
materia. Para esto tendremos en cuenta que en equilibrio $\beta$ la
fracción de electrones es despreciable, también consideraremos que
las masas de los quarks $u,d$ son nulas. En este caso tenemos:

\begin{equation}\label{ec de ud masless}
    n_f=\frac{\mu_f^3}{\pi^2} \ \ \ \
    \epsilon_f=\frac{3\mu_f^4}{4\pi^2} \ \ \ \
    P_f=\frac{\mu_f^4}{4\pi^2},
\end{equation}

Utilizando el modelo de Bag del MIT e imponiendo la condición de
equilibrio $(P=0)$ obtenemos   para la energía y la presión total
dentro del bag~(\ref{pres&energia_bag1})
y~(\ref{pres&energia_bag2}):

\begin{eqnarray}
   B_{bag} &=& \sum_fP_f \\
   \epsilon &=& 4B_{bag},
\end{eqnarray}

La neutralidad de carga en este caso toma la forma:
$n_d=2n_u\,\,\,\Rightarrow\,\,\,\mu_u=2^{-1/3}\mu_d\equiv\mu_{ud}$,
para la presión tenemos:
$P_{ud}=P_u+P_d=(\frac{1+2^{4/3}}{4\pi^2})\mu_{ud}^4=B_{bag}$ y la
conservación del número bariónico nos da:
$n_{B\,ud}=\frac{n_u+n_d}{3}=\mu_{ud}/\pi^2$, con estas ecuaciones
podemos calcular la energía por barión para el gas formado por
quarks $u$ y $d$:

\begin{equation}\label{EporB_ud}
\frac{\epsilon_{ud}}{n_{B\,ud}}=(1+2^{4/3})^{1/4}(4\pi^2)^{1/4}B_{bag}^{1/4}=6.411B_{bag}^{1/4}\simeq943\,\mathrm{MeV}
\end{equation}

\noindent donde hemos tomado $B_{bag}^{1/4}=145\,\mathrm{MeV}
\,\,\,\Rightarrow\,\,\,B_{bag}\simeq57\,\mathrm{MeV}\,\,\mathrm{fm}^{-3}$.
La energía por barión para un gas de neutrones es la masa del
neutrón $(m_n=939,6\,\mathrm{MeV})$, para un gas de
$^{56}\mathrm{Fe}$ la energía por barión se puede calcular como:
$E/A|_{^{56}\mathrm{Fe}}=(56m_N-56\cdot8.8)/56=930\,\mathrm{MeV}$.
Por tanto la estabilidad del gas formado por quarks $u$ y $d$ con
respecto al gas de neutrones requiere que
$\epsilon_{ud}/n_{B\,ud}<m_n\,\,\,\Rightarrow\,\,\,B_{bag}<60\,\mathrm{MeV}\,\mathrm{fm}^{-3}
$ y con respecto al gas de $^{56}\mathrm{Fe}$ que
$\epsilon_{ud}/n_{B\,ud}<E/A|_{^{56}\mathrm{Fe}}\,\,\,\Rightarrow\,\,\,B_{bag}<57\,\mathrm{MeV}\,\mathrm{fm}^{-3}$.

En la naturaleza se observan los neutrones y el $^{56}\mathrm{Fe}$
pero no el gas de quarks $u$ y $d$, por lo que se concluye que el
$B_{bag}$ debe ser mayor que los valores antes mencionados. Este
hecho hace que se tome el valor $B_{bag}=57 \,\mathrm{MeV\,fm}^{-3}$
como un límite inferior para este parámetro.


\subsection{Estabilidad de la Materia Extraña de Quarks en equilibrio estelar.}

Analicemos ahora la estabilidad de la MEQ, es decir, materia formada
por quarks $u$, $d$ y $s$ libres. El equilibrio químico de este
sistema está garantizado por las interacciones débiles:
\begin{eqnarray*}
  d &\leftrightarrow& u+e^-+\bar{\nu_e} \\
  s &\leftrightarrow& u+e^-+\bar{\nu_e} \\
  u+s &\leftrightarrow& d+u
\end{eqnarray*}
estas reacciones conducen a relaciones entre los potenciales
químicos. Tomaremos igualmente que $\mu_{\bar{\nu}_e}=0$ pues los
neutrinos escapan de la estrella. El equilibrio $\beta$, la
neutralidad de carga y la conservación del número bariónico quedan
expresados mediante:
\begin{subequations}\label{consNb}
\begin{eqnarray}
\mu_u+\mu_e-\mu_d=0 \,\,\,, \,\,\, \mu_d-\mu_s&=&0,\\
2n_u-n_d-n_s-3n_e&=&0, \\
n_u+n_d+n_s-3 n_{B}&=&0.
\end{eqnarray}
\end{subequations}

Al igual que en el caso de la materia de quarks normal supondremos
que la fracción de electrones es muy pequeña, por tanto la
neutralidad de carga queda como: $n_u=n_d=n_s
\,\,\,\Rightarrow\,\,\,\mu_{uds}=\mu_u=\mu_d=\mu_s$. Para un
$B_{bag}$ fijo el gas de $u\,,d\,,s$ ejerce la misma presión que el
gas de $u$ y $d$, también la energía es la misma
$(\epsilon_{uds}=4B_{bag})$. Esto ocurre cuando
$\mu_{uds}=3^{-1/4}(1+2^{4/3})^{1/4}\mu_{ud}$ con lo que se obtiene
el número bariónico:
$n_{B\,uds}=\mu_{uds}/\pi^2=[(1+2^{4/3})/3]^{3/4}n_{B\,ud}$, esto
nos da una energía por barión:
\begin{equation}\label{eporB_uds}
    \frac{\epsilon_{uds}}{n_{B\,uds}}=3^{3/4}(4\pi^2)^{1/4}B_{bag}^{1/4}=5.71B_{bag}^{1/4}\simeq829\,\mathrm{MeV},
\end{equation}
este valor es 100 veces menor que para la materia normal de quarks
(para un mismo valor de $B_{bag}$), por lo que podemos decir que se
cumplen las siguientes desigualdades:
\begin{equation}\label{stabineq_uds}
\left.\frac{E}{A}\right|_{MEQ}<\left.\frac{E}{A}\right|_{^{56}\mathrm{Fe}}
<\left.\frac{E}{A}\right|_{u,d},
\end{equation}

donde el parámetro $B_{bag}$ se encuentra entre $57-90\,MeV
fm^{-3}$. Este resultado conduce directamente a la hipótesis de
Bodmer--Witten para el caso de ambientes ricos en quarks $s$ como
puede ser el interior de las ENs~\cite{Schmitt:2010pn}.

En el capítulo siguiente estudiaremos los efectos del campo
magnético en la estabilidad de la MEQ y de la materia normal de
quarks.


\chapter{Propiedades termodinámicas de la Materia Extraña de Quarks.}\label{cap3}
\markright{Capítulo 3: Propiedades termodinámicas de la Materia
Extraña de Quarks.}

\noindent

\section{Materia extraña de quarks a $B=0$ y $T=0$.}

De forma general el gran potencial termodinámico para el gas de
quarks relativistas y libres, en el marco del modelo de Bag, viene
dado por cuatro términos~\cite{Chakrabarty:1996te}:
\begin{equation}\label{OTotal}
\Omega=B_{bag}V+\sum_{i}\left[
\Omega_{i,v}(\mu_{i},T)V+\Omega_{i,s}(\mu_{i},T)S+\Omega_{i,c}(\mu_{i},T)C
\right],
\end{equation}
donde la suma se extiende sobre $i=(u,d,s,e)$, y tiene en cuenta
todas las especies de fermiones presentes: electrones, quarks $u$,
$d$ y $s$; $\mu_{i}$ es el potencial químico de cada gas de
partículas; $T$ la temperatura; $V,S$ y $C$ son el volumen dentro
del cual se encuentra el gas, el área y curvatura de la superficie
del Bag respectivamente.

\begin{figure}[h!t]\label{modelo_B0}
\begin{center}
\includegraphics[height=3.0in,width=6.0in]{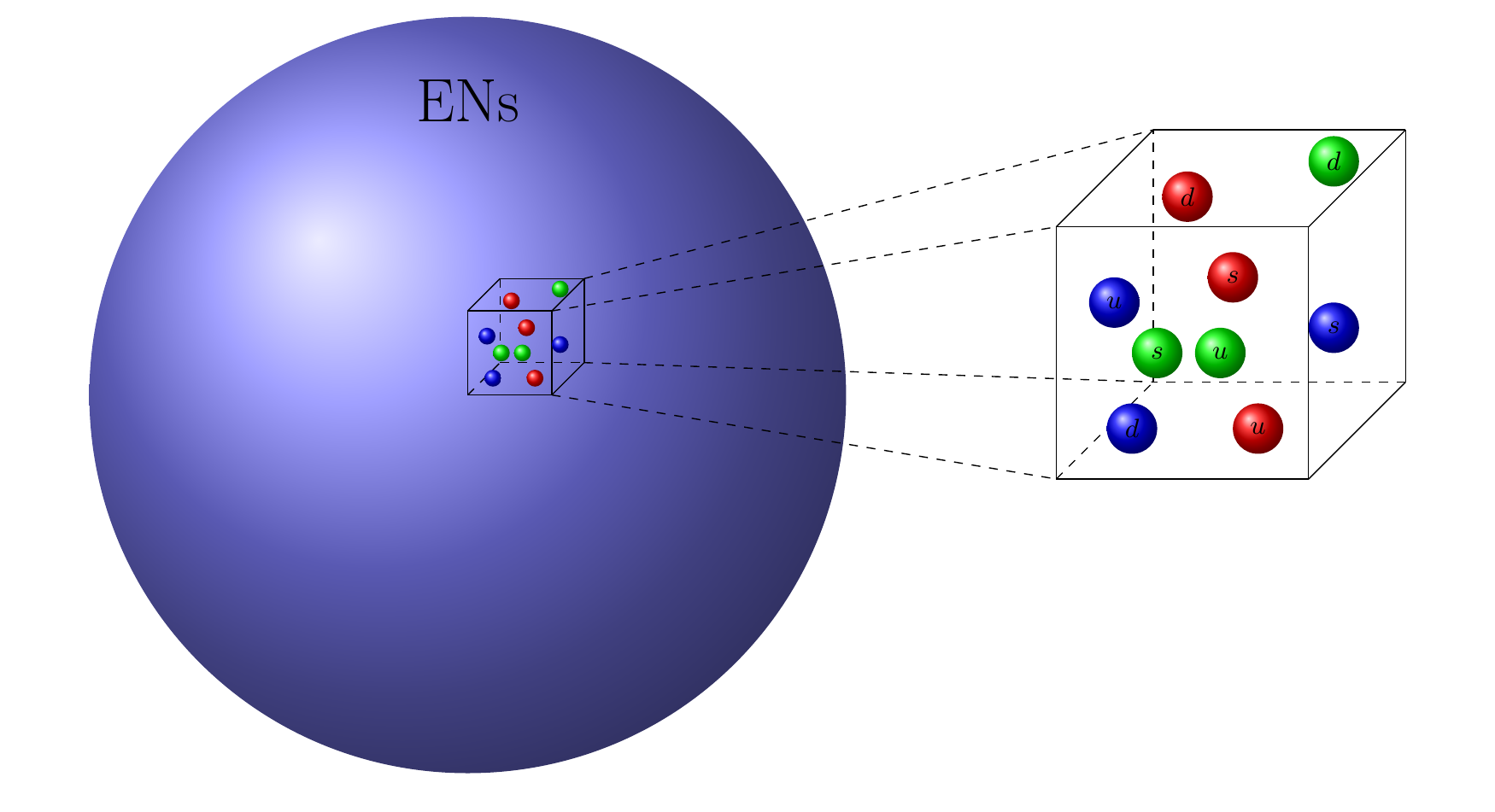}
\end{center}
\caption{Volumen local en el cual los quarks se comportan como
partículas libres.}
\end{figure}

Consideraremos en nuestro estudio un volumen en el interior de la
estrella por lo que no tendremos en cuenta los efectos superficiales
ni de curvatura. Dentro de este volumen, los quarks $u, \, d, \, s$
se comportan como partículas libres \fig{modelo_B0}.
El término volumétrico del Gran Potencial Termodinámico para
fermiones libres está dado por la expresión~\cite{Landau PHE}:
\begin{equation}\label{OVol}
\Omega_{i,v}(\mu_{i},T)=-\dfrac{ d_{i}T}{(2\pi)^{3}}\int_{p}[\ln
[f^{+}(p,\mu_{i},T)f^{-}(p,\mu_{i},T)]d^{3}p],
\end{equation}
\begin{equation}\label{FDDistrib}
f^{\pm}(p,\mu_{i},T)=1+\exp\{-\dfrac{E_{p,i}\mp\mu_{i}}{T}\},
\end{equation}
aquí $f^{\pm}$ representa las contribuciones de las partículas y
antipartículas ($f^{+}$ para partículas y $f^{-}$ para
antipartículas), $E_{p,i}$ es la energía de las partículas dada por
el espectro y $d_{i}$ es el factor de degeneración: $1$ para
electrones y $3$ para quarks\footnote{Se toma por convenio las
unidades naturales $\hbar=c=k=1$.}. El Gran Potencial juega un papel
importante ya que de él se derivan todas las propiedades
termodinámicas del sistema~\cite{Landau PHE,CAMAYTE} y se pueden
obtener las EdE del sistema.

En~(\ref{OVol}) y~(\ref{FDDistrib}), $E_{p,i}$ se obtiene de la
solución de la ecuación de Dirac para fermiones libres:
\begin{equation}\label{DiracB0}
    E_{p,i}=\sqrt{p^2+m_i^2},
\end{equation}
donde $\vec{p}$ y $m_{i}$ representan el momentum lineal y la masa
en reposo de cada partícula respectivamente.

En nuestro caso podemos hacer un cambio de variables a coordenadas
esféricas e integrar por partes con lo que obtenemos:
\begin{equation}\label{OmegaT01}
\Omega_{i,v}(\mu_{i},T)=-\dfrac{
d_{i}}{6\pi^{2}}\int_{0}^{\infty}\frac{p^4\,dp}{\sqrt{p^2+m^2}}\Bigl(n_F(\frac{E_{p,i}-\mu_i}{T})+n_F(\frac{E_{p,i}+\mu_i}{T})\Bigr),
\end{equation}
donde $n_F(x)=1/(e^x+1)$ es la función de distribución de
Fermi--Dirac.

En el caso de una ENs o una EQs la temperatura es mucho menor que la
temperatura de Fermi $(T/T_F\sim10^{-4})$, por lo que podemos
considerar el límite degenerado $(T=0)$. Esto hace que
$n_F=\theta(\mu_i-E_{p,i})$ donde  $\theta(x)$ es la función paso
unitario, además la contribución de las antipartículas es nula pues
$n_F(\frac{E_{p,i}+\mu_i}{T})\xrightarrow{\  T\to0 \ }0$ con lo que
obtenemos:
\begin{equation}\label{OmegaT02}
\Omega_{i,v}(\mu_{i},0)=-\dfrac{
d_{i}}{6\pi^{2}}\int\limits_{0}^{\sqrt{\mu_i^2-m^2}}\frac{p^4\,dp}{\sqrt{p^2+m^2}}.
\end{equation}

La integral~(\ref{OmegaT02}) da como resultado:
\begin{equation}\label{OmegaT0}
\Omega_{i,v}(\mu_{i},0)=-\Omega_{0}\{x_{i}(x_{i}^{2}-\frac{5}{2})p_{F,i}+\dfrac{3}{2}\cosh^{-1}[x_{i}]\},
\end{equation}
donde hemos definido la variable adimensional $x_i=\mu_i/m_i$,
$\Omega_{0}=\dfrac{d_{i}m_{i}^{4}}{24\pi^{2}}$ y
$p_{F,i}=\sqrt{x_{i}^{2}-1}$ es el momentum de Fermi adimensional de
cada especie.

A partir de~(\ref{OmegaT0}) podemos obtener la presión, la densidad
de partículas y la energía del sistema:
\begin{eqnarray}
  P            &=& -\sum_{i}\Omega_{i,v} \\ \label{Pres}
  N_{i}        &=& -\dfrac{\partial\Omega_{i,v}}{\partial\mu_{i}} \\
  \label{densidad}
  \epsilon_{i} &=& \Omega_{i,v}+\mu_{i}N_{i},\label{energia}
\end{eqnarray}
la energía total viene dada por la
ecuación~(\ref{pres&energia_bag2}). Se recuperan las
expresiones~(\ref{ec de ud masless}) cuando tomamos $m_i=0$.

\section{Materia Extraña de Quarks Magnetizada a $T=0$.}

\begin{figure}[h!t]\label{modelo_B}
\begin{center}
\includegraphics[height=3.0in,width=6.0in]{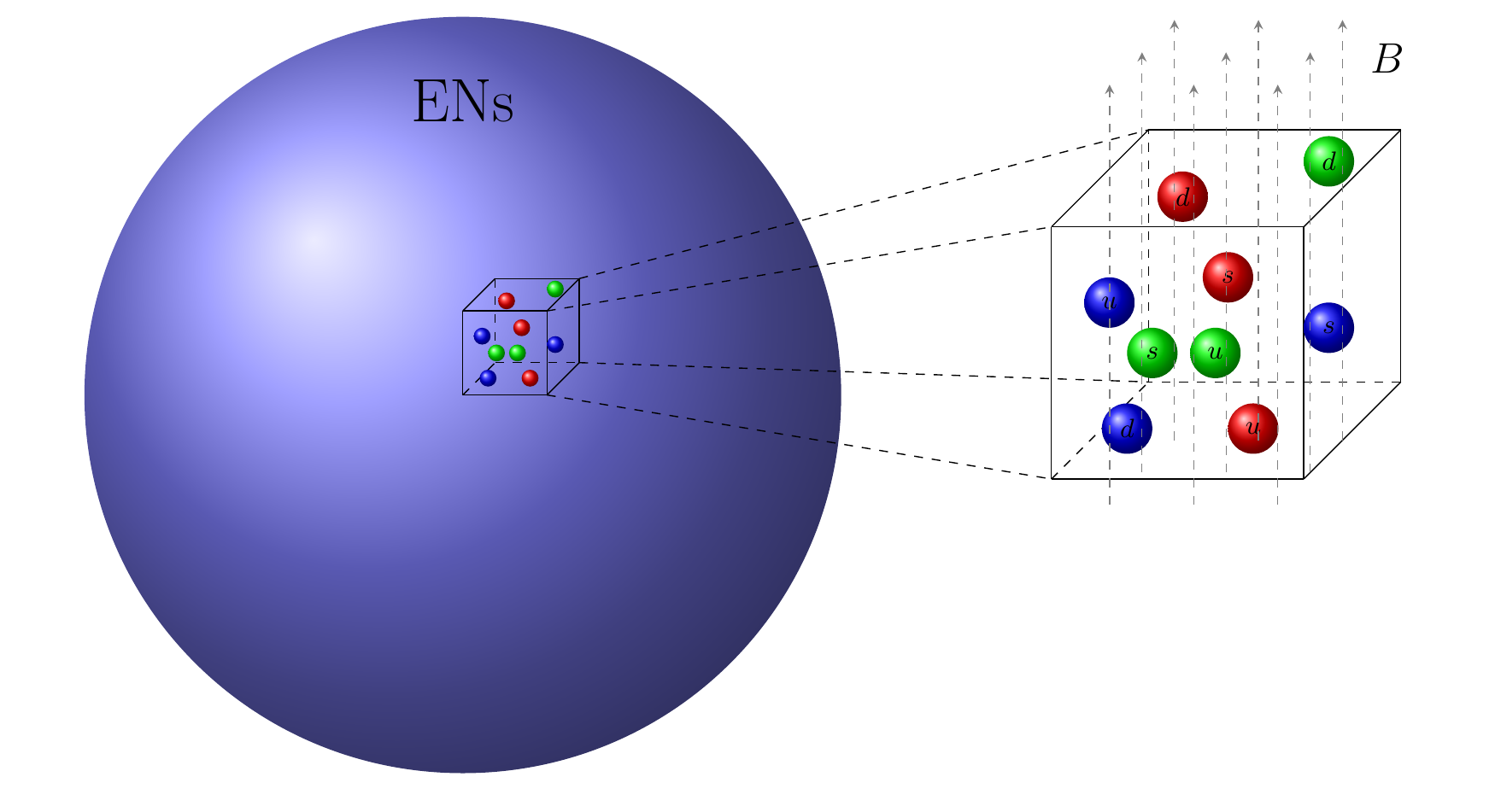}
\end{center}
\caption{Igual que en la \fig{modelo_B0}, pero considerando un campo
magnético constante y homogéneo.}
\end{figure}

En el caso de la MEQM, estudiaremos igualmente un volumen local con
un campo magnético, que consideraremos constante, homogéneo y en la
dirección del eje $z$ \fig{modelo_B}.

El espectro energético se obtiene al resolver la ecuación de Dirac
para fermiones en presencia de un campo magnético y tiene la
forma~\cite{Bagrov}:
\begin{equation}\label{espectroquarks}
E_{p,i}^{n,\eta}=\sqrt{p_{z}^{2}+p_{\perp}^{2}+m_{i}^{2}},\;\;p_{\perp}^{2}=m_{i}^{2}\{(\sqrt{\dfrac{B}{B_{i}^{c}}(2n+1-\eta)+1}-\eta
y_{i}B)^{2}-1\}.
\end{equation}

$$\;\;\;B_{i}^{c}=\dfrac{m_{i}^{2}}{q_{i}},\;\;y_{i}=\dfrac{Q_{i}}{m_{i}},\;\;\tau=\dfrac{\alpha}{2\pi}\dfrac{e}{2m_{e}}.$$
donde $n$ son los niveles discretos de Landau, $\eta=\pm1$ son los
autovalores del operador de spin, $B$ es el campo magnético,
$B_{i}^{c}$ es el valor del campo crítico (campo para el cual la
energía ciclotrónica se iguala a la energía en reposo), $q_{i}$,
$Q_{i}$ y $m_{i}$ representan la carga eléctrica, el momento
magnético anómalo (MMA) y la masa de cada partícula respectivamente.
La carga eléctrica de cada partícula está en la \tab{partículas},
los valores de MMA y la masa que utilizaremos se dan en la
\tab{tabla1}.

\begin{table}[!ht]
\begin{center}
\label{tabla1}
\begin{tabular}{|c|c|c|c|c|}
\hline
\textbf{Partículas} & $u$ & $d$ & $s$ & $e$ \\
\hline
$Q_{i}$&$1.85\mu_{N}^{\,\,\dag}$&$-0.97\mu_{N}$&$-0.58\mu_{N}$&$1.16\times10^{-3}\mu_{B}^{\,\,\ddag}$\\
\hline
$m_{i}\;(\mathrm{MeV})$&5&5&150&0.5\\
\hline

\multicolumn{5}{l}{\small$^{\dag} \text{Magneton Nuclear}\,\,\mu_{N}=e/2m_{p}\simeq3.15\times10^{-18}\mathrm{MeV}\;\mathrm{G}^{-1}$.} \\
\multicolumn{5}{l}{\small$^{\ddag} \text{Magneton de Bohr}\,\,
\mu_{B}=e/2m_{e}\simeq5.79\times10^{-15}\mathrm{MeV}\,\mathrm{G}^{-1}$.}

\end{tabular}\caption{Valores del MMA y masa de las
partículas consideradas en nuestro estudio.}
\end{center}
\end{table}

Análogamente al proceso que seguimos para calcular el potencial
termodinámico para el gas de quarks sin campo magnético, en el caso
de campo tenemos que calcular la integral~(\ref{OVol}).

En presencia del campo magnético la simetría deja de ser esférica,
por lo que para integrar realizamos un cambio a coordenadas
cilíndricas obteniendo:
\begin{equation}
\Omega_{i,v}(\mu_{i},T)=-\dfrac{
d_{i}T}{(2\pi)^{3}}\sum_{\eta=\pm1}\sum_{n=1}^{+\infty}\int_{-\infty}^{+\infty}[\ln[1+\exp\{-\dfrac{E_{p,i}^{n,\eta}-\mu_{i}}{T}\}]dp_{z}],
\end{equation}
\begin{equation}
\int_{-\infty}^{+\infty}\int_{-\infty}^{+\infty}[dp_{x}dp_{y}]=2\pi
q_{i}B.
\end{equation}

El potencial termodinámico tiene la forma:
\begin{equation}\label{Om_mag}
   \Omega =-B\sum_i {\mathcal M}^0_{i}\sum_n^{n_{max}^i}\sum_{\eta=\pm1}\left ( x_{i}p_{F,i}^{\eta} -
   h^{\eta\,2}_{i}\ln\frac{x_{i} +
 p^{\eta}_{F,i}}{h^{\eta}_{i}}\right ),
\end{equation}
donde
\begin{eqnarray} \label{dimvar}
{\mathcal M}^0_{i}&=& \frac{e_i d_i m_{i}^2}{4\pi ^2}, \quad p_{F,i}^{\eta}=\sqrt{x_{i}^2-h_{i}^{\eta}\;^2},\\
h_{i}^{\eta} &=&\sqrt{\frac{B}{B^{c}_i}\, (2n + 1-\eta) +1}-\eta
y_{i}B.
\end{eqnarray}
En la ecuacion~(\ref{Om_mag}) la suma sobre los niveles de Landau
$n$ es hasta el valor entero $n_{max}^i = I\left[\left((x_{i} + \eta
y_iB)^2 -1\right)\,B^{c}_i/(2B)\right]$, $p_{F,i}$ y $h_{i}^{\eta}$
son el momento de Fermi y la masa magnética respectivamente.

La presencia del campo magnético hace que surja una anisotropía en
la presión, de tal forma que la podamos dividir en paralela
$P_{\parallel}$ y perpendicular al campo $P_{\perp}$, por lo que el
tensor energía impulso~(\ref{Tensor EM}) queda:
\begin{equation}\label{Tensor_EM_mag}
\mathcal{T}^{\mu}_{\,\,\,\,\nu}=diag(\epsilon,P_{\perp},P_{\perp},P_{\parallel}),
\end{equation}
donde $P_{\parallel}=-\Omega$ y $P_{\perp}=-\Omega-{\mathcal M}B$
siendo ${\mathcal M}=-\partial\Omega_{i,v}/\partial B$ la
magnetización.

Los parámetros termodinámicos se expresan mediante las expresiones:
\begin{equation}
N= \sum_i N^0_{i}\frac{B}{B^c_{i}}\sum_n^{n_{max}^i}\sum_{\eta=\pm1}
p^{\eta}_{F,i}\,, \label{TQf}
\end{equation}
\begin{equation}
\varepsilon=\Omega +\mu N =B\sum_i{\mathcal
M}^0_{i}\sum_n^{n_{max}^i}\sum_{\eta=\pm1}\left (
x_ip^{\eta}_{F,i}+h_{i}^{\eta\,2}\ln\frac{x_i+p_{F,i}^{\eta}}{h_{i}^{\eta}}\right
),\label{TQi}
\end{equation}
\begin{equation} \label{Ppar}
  P_{\parallel}=-\Omega =B\sum_i {\mathcal M}^0_{i}\sum_n^{n_{max}^i}\sum_{\eta=\pm1}\left ( x_{i}p_{F,i}^{\eta} -
   h^{\eta\,2}_{i}\ln\frac{x_{i} +
 p^{\eta}_{F,i}}{h^{\eta}_{i}}\right ),
\end{equation}
\begin{equation}\label{Pper}
 P_{\perp} = -\Omega -{\mathcal M} B=B\sum_i {\mathcal M}^0_{i} \sum_n^{n_{max}^i}\sum_{\eta=\pm1}\left (2h_{i}^{\eta}\gamma_i^{\eta}\ln\frac{x_{i} + p_{F,i}^{\eta}}{h_{i}^{\eta}}\right),
\end{equation}
donde
\begin{eqnarray} \label{dimvar2}
\quad N_{i}^0 &=& \frac{d_i m_{i}^3}{2\pi^2},\\
\gamma^{\eta}_i&=&\frac{B\,(2n+1-\eta)}{2B^c_i\sqrt{(2n+1-\eta)B/B^c_i+1}}-\eta
y_i B,
\end{eqnarray}
son magnitudes adimensionales.

Las expresiones anteriores contienen las contribuciones del
diamagnetismo de Landau (cuantización de los niveles de Landau), y
el paramagnetismo de Pauli debido a la presencia del MMA. El
paramagnetismo de Pauli no modifica las EdE ni la estabilidad de la
MEQM, por lo que no tendremos en cuenta su efecto en lo que sigue.

\subsection{Ecuaciones de estado para la MEQ.}

\begin{figure}[h!t]
      \centering
      \includegraphics[height=8cm,width=9cm]{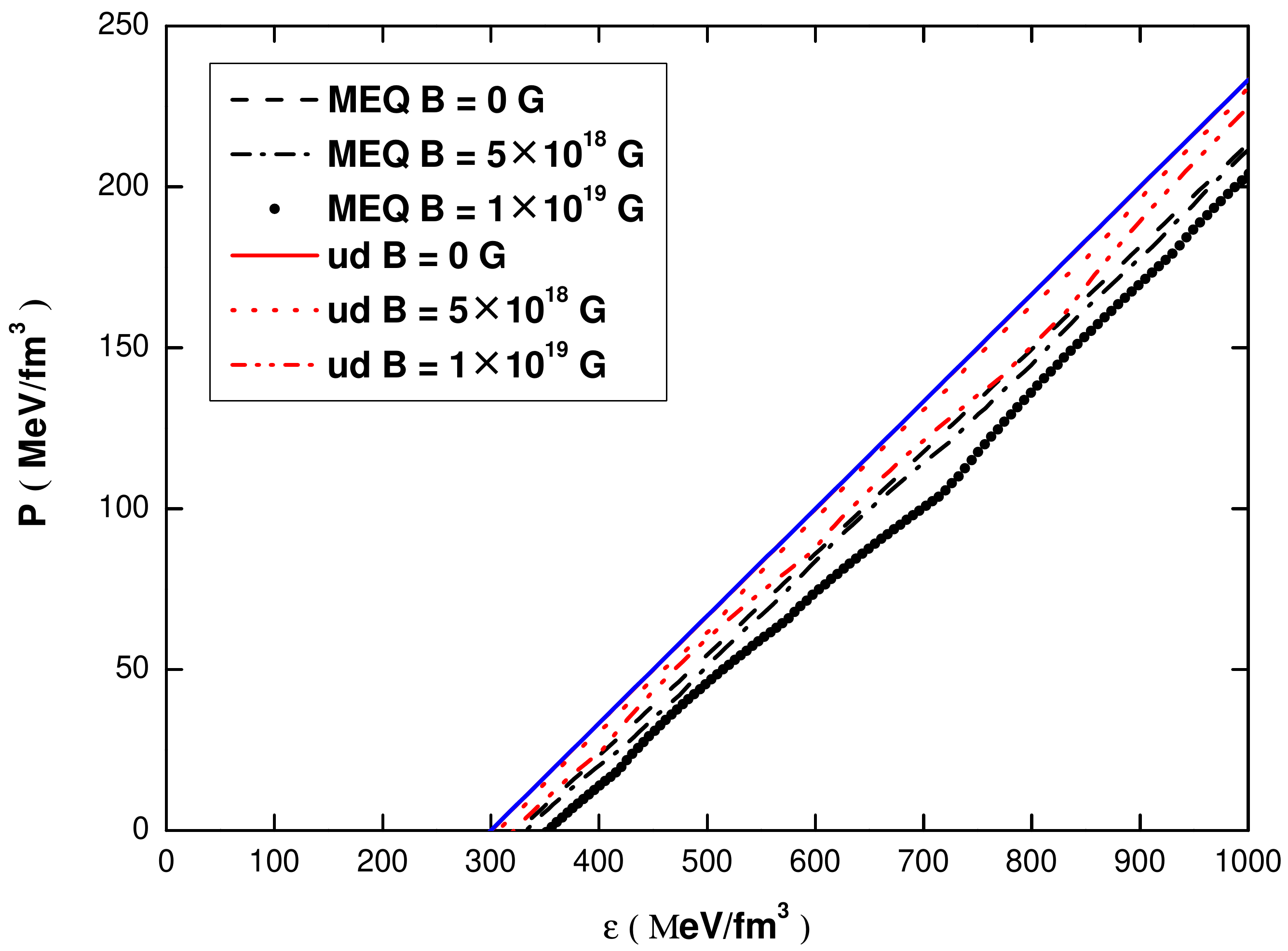}
      \caption{Ecuación de estado para la MEQ: presión contra energía para diferentes valores de campo.
      Para comparar se han añadido las gráficas para la materia normal en equilibrio con electrones.}\label{EoS}
\end{figure}

Para determinar las EdE de la MEQM en equilibrio estelar, hay que
resolver numéricamente las ecuaciones~(\ref{consNb}). En la
\fig{EoS} mostramos las EdE para la MEQ $(B=0)$ y para la MEQM para
$B=5\times 10^{18}\,\mathrm{G}$, en todos los casos hemos tomado el
valor $B_{bag}=75\, \mathrm{MeV\,fm^{-3}}$. Para comparar hemos
incluido las EdE de la materia normal ($u$, $d$ en equilibrio con
electrones).

El caso $B=0$ no muestra diferencias con la ecuación de estado
$P=(\epsilon -4B_{bag})/3$ de la materia formada por quarks $u$ y
$d$ sin masa, analizada en el epígrafe~(\ref{estmatud}), lo que
muestra que considerar las masas de los quark $u$ y $d$ diferentes
de cero, siendo ellas tan pequeñas, no influye de manera
significativa en los resultados.

Aunque las variaciones que el campo magnético produce en las EdE no
son muy notables, ellas pueden alterar las propiedades macroscópicas
como son la masa y el radio de la estrella como veremos en  el
siguiente capítulo.

\section{Estabilidad de la materia de quarks magnetizada.}

Estudiemos la estabilidad de la  MEQM como función de los parámetros
del modelo, es decir, la densidad bariónica $(n_B)$, el campo
magnético, el parámetro de Bag y la masa del quark $s$ $m_s$. Debido
a la presencia del campo magnético, la anisotropía de las presiones
implica que $P_{\perp}<P_{\parallel}\,$~\cite{Felipe:2007vb}. La
condición de estabilidad se expresa mediante la relación:
\begin{equation} \label{stabpper}
P_{\perp} =\sum_{i}P_{\perp,i}- B_{\rm bag}=0\,.
\end{equation}

Para investigar la estabilidad de la MEQM en equilibrio estelar
resolveremos el sistema de ecuaciones~(\ref{consNb})
y~(\ref{stabpper}), utilizando los resultados obtenidos
en~(\ref{Pper})--~(\ref{TQf}). Al resolver este sistema de
ecuaciones se obtiene la región en la cual los parámetros cumplen
con las desigualdades de estabilidad:
\begin{equation}\label{stabineq}
\left.\frac{E}{A}\right|_{MEQ}^B<\left.\frac{E}{A}\right|_{MEQ}^{B=0}<\left.\frac{E}{A}\right|_{^{56}\text{Fe}}
<\left.\frac{E}{A}\right|_{u,d}^{B}<
\left.\frac{E}{A}\right|_{u,d}^{B=0}.
\end{equation}

\begin{figure}[h!t]
\centering
\begin{minipage}{8cm}
\includegraphics[height=7cm,width=8cm]{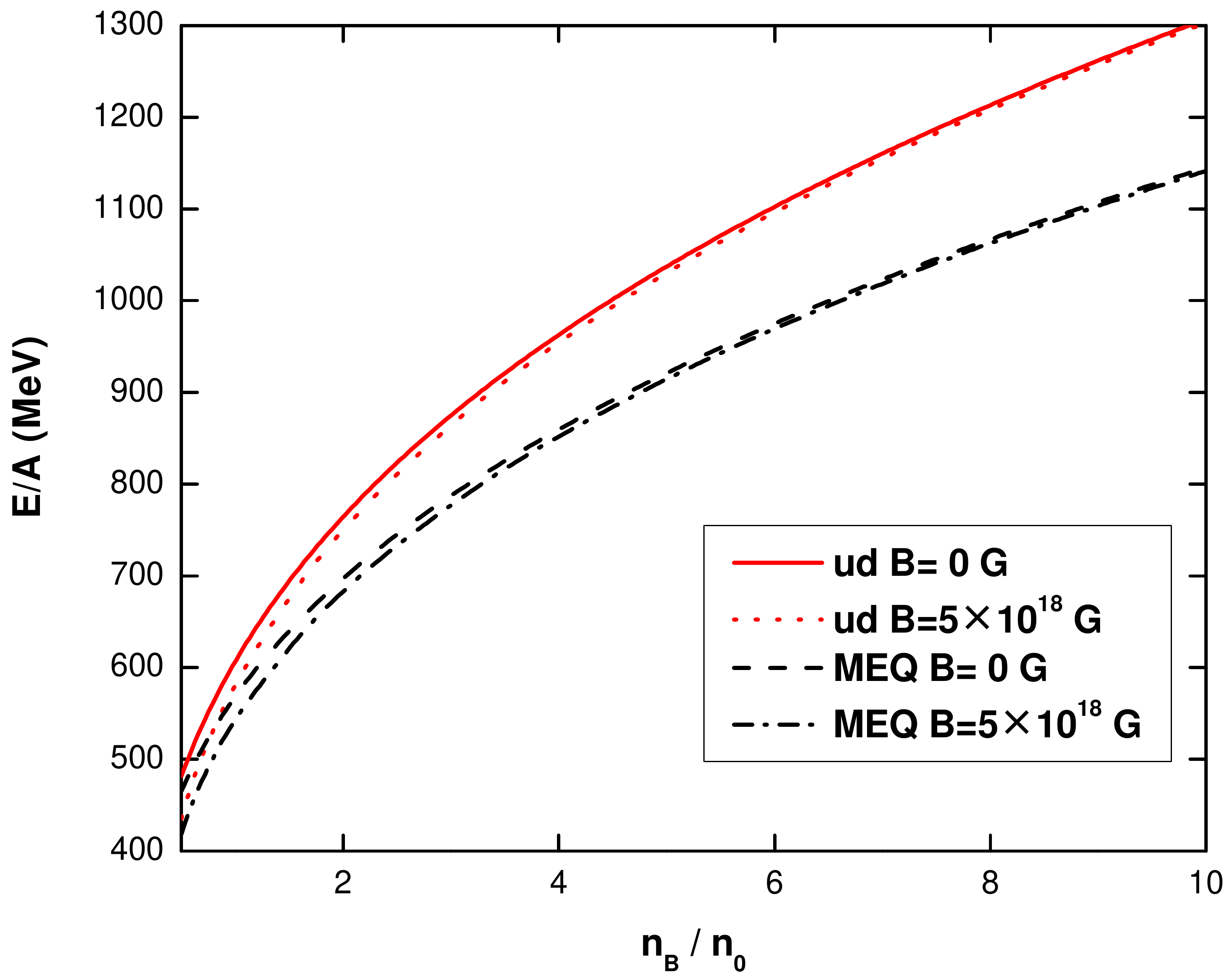}
      \caption{\small Energía por barión en función de la densidad
      bariónica para $B=0$ y $B=5\times 10^{18}$~G, asumiendo la
      condición de estabilidad de presión cero.}\label{Energy}
\end{minipage}
\hfill
\begin{minipage}{8cm}
 \includegraphics[height=7cm,width=8cm]{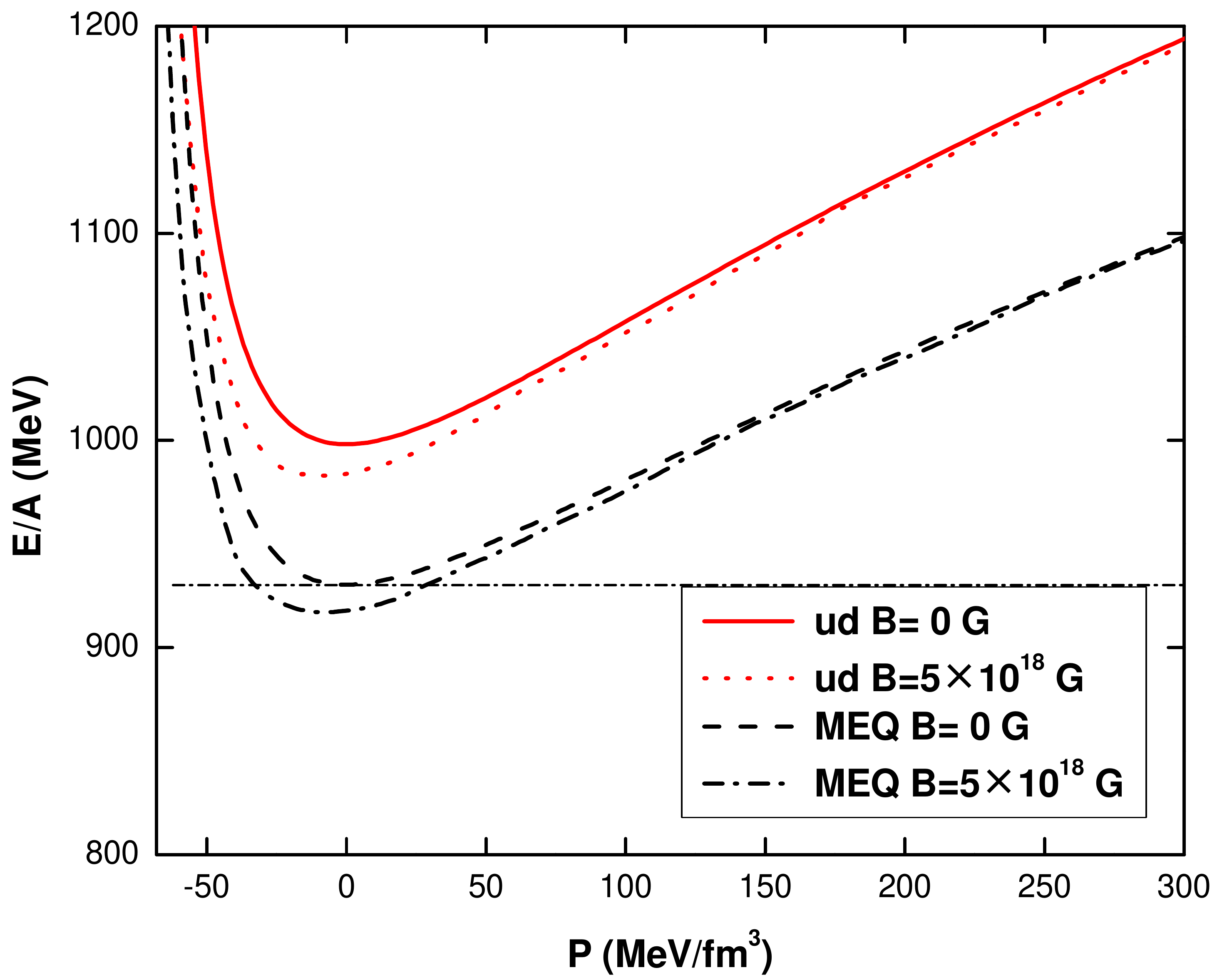}
      \caption{\small Energía por barión contra presión para $B=0$ y $B=5\times
      10^{18}\,\mathrm{G}$. Tomando $B_{\rm bag}=75\,\mathrm{MeV\,fm}^{-3}$. La linea
      horizontal de puntos corresponde a $\left. E/A\right|_{^{56}\text{Fe}} \simeq 930\,\mathrm{MeV}$.}\label{EP}
\end{minipage}
\end{figure}
En la \fig{Energy} presentamos una comparación de la energía por
barión $E/A$ (equivalentemente $\varepsilon/n_{B}$) contra la
densidad de partículas $n_B/n_0$ para $B=0$ y $B=5\times 10^{18}$~G
exigiendo la condición de equilibrio $P_{\perp}=0$. Hemos asumido
que $m_u = m_d = 5$~MeV y $m_s =150$~MeV en todos los casos.

Como puede observarse en la figura $E/A$ es menor cuando está
presente el campo magnético. Para $B=5\times 10^{18}$~G obtenemos
que $E/A\approxeq 919\,\mathrm{MeV}$, $n_B/n_0\approxeq 2.2$ y
$B_{bag}\approxeq 75\,\mathrm{MeV\,fm}^{-3}$ mientras que para $B=0$
e igual valor del Bag, $E/A\approxeq 929\,\mathrm{MeV}$,
$n_B/n_0\approxeq 2.1$.

El comportamiento de $E/A$ con la presión se muestra en la \fig{EP}
fijando $B_{bag}=75\,\mathrm{MeV\,fm}^{-3}$. Podemos notar que el
punto de presión cero para la MEQM se alcanza para un valor de
densidad de energía menor que en el caso de la MEQ. Por tanto la MEQ
es más estable y más compacta en presencia de campo magnético.
\begin{figure}[h!t]
\centering
\begin{minipage}{8cm}
\includegraphics[height=7cm,width=8cm]{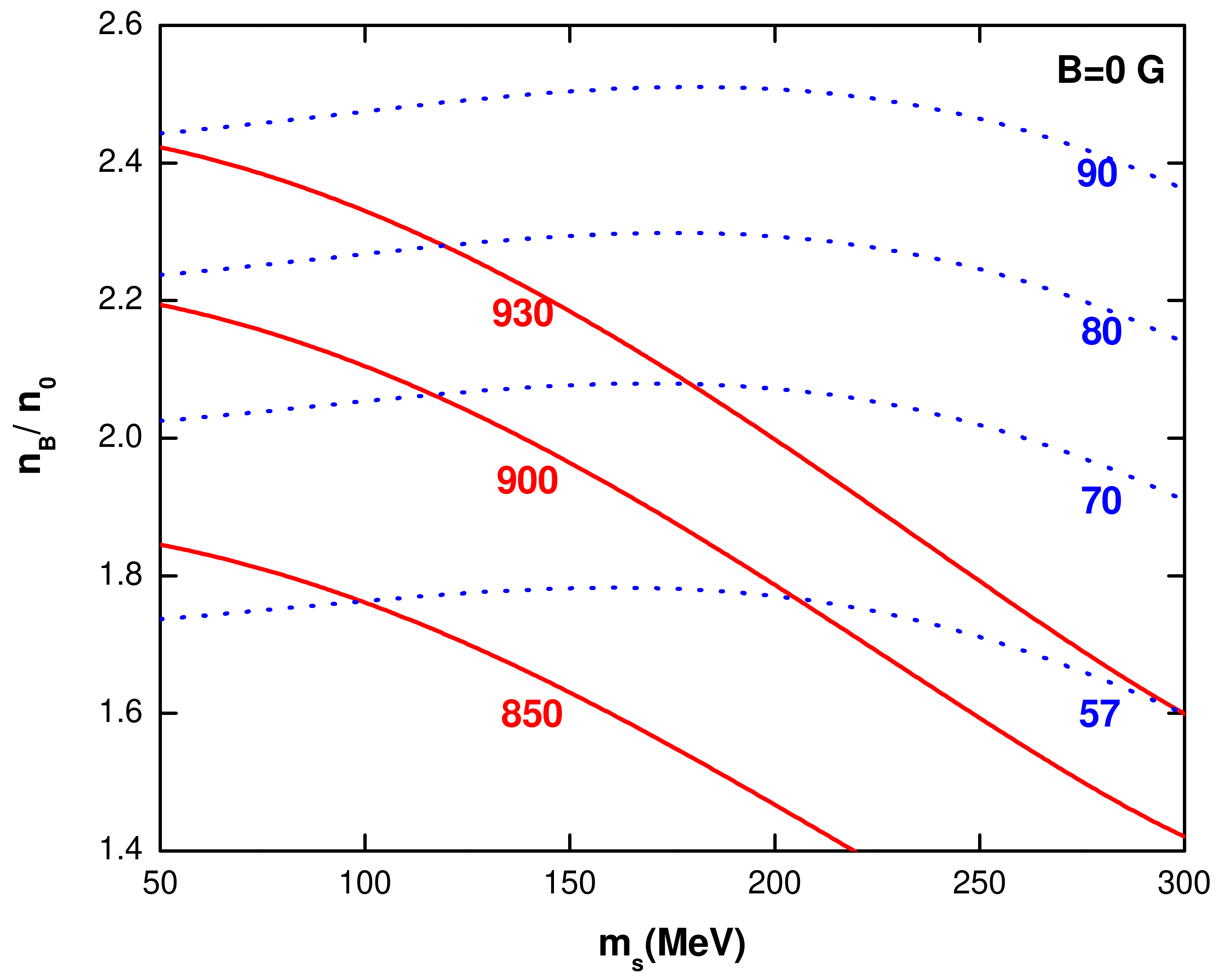}
      \caption{\small Ventanas de estabilidad para la MEQ en el plano
      ($m_s,n_B$). Se muestran los contornos de $B_{bag}=const$ y $E/A=const$.}\label{cBag}
\end{minipage}
\hfill
\begin{minipage}{8cm}
 \includegraphics[height=7cm,width=8cm]{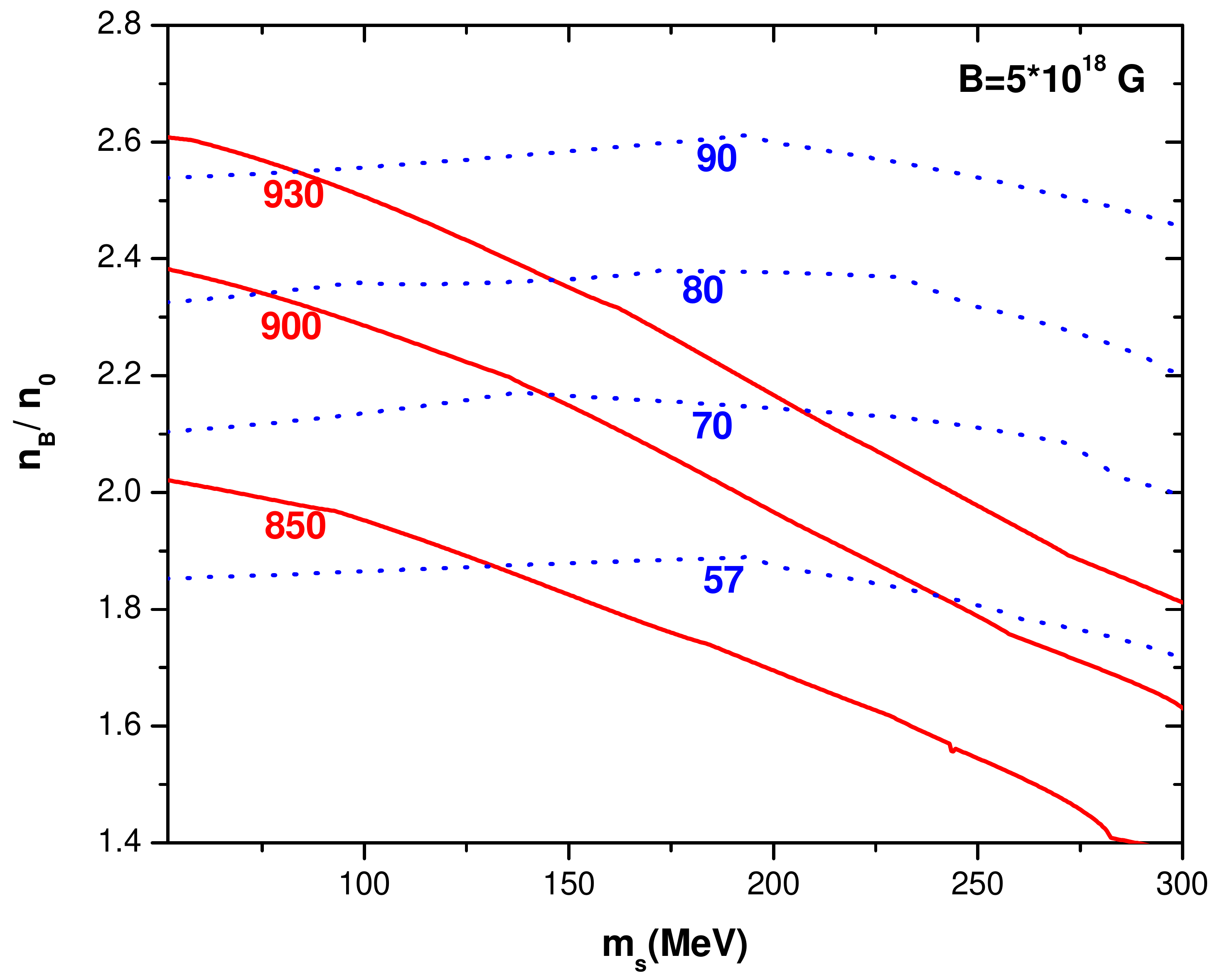}
      \caption{\small Ventanas de estabilidad para la MEQM en el plano
      ($m_s,n_B$). Se muestran los contornos de $B_{bag}=const$ y $E/A=const$.}\label{cE}
\end{minipage}
\end{figure}

Un estudio más detallado de la estabilidad lo podemos realizar a
través de las ventanas de estabilidad, las cuales nos brindan una
mayor información acerca de los límites para los parámetros entre
los cuales se cumple la desigualdad de estabilidad~(\ref{stabineq}).
Para investigar cómo el campo magnético afecta estas ventanas,
consideraremos las regiones de estabilidad en el plano ($m_s,n_B$).
Para ilustrar, fijaremos el campo magnético en el valor $B=5\times
10^{18}\,\mathrm{G}$, $B=0\,\mathrm{G}$ y estudiaremos los contornos
$B_{bag}=const$ y $E/A=const$.

Estos resultados se presentan en las \fig{cBag} y~\fig{cE} para MEQ
y MEQM respectivamente. Como puede observarse, el campo magnético
hace que las ventanas de estabilidad para la MEQ se muevan hacia
valores mayores de la densidad bariónica. Además se modifica el
intervalo permitido de valores para el parámetro de Bag debido a que
por debajo del contorno de energía de 930 MeV las EdE para MEQM
corresponden a un valor de $E/A$ a presión cero menor que el del
$^{56}$Fe
($\left.\frac{E}{A}\right|_{MEQ}^B<\left.\frac{E}{A}\right|_{^{56}\text{Fe}}$).

Del análisis de la \fig{cE} podemos considerar la MEQM como
absolutamente estable si $57\,\mathrm{MeV\,fm}^{-3}\, <B_{bag}<\,
90\,\mathrm{MeV\,fm}^{-3}$, $1.85\,<n_B/n_0<\,2.6$ y para un amplio
rango de masas del quark $s$, en particular para
$m_s=150\,\mathrm{MeV}$ y $B_{bag}=75\,\mathrm{MeV\,fm}^{-3}$
valores que utilizaremos en el próximo capítulo para calcular
algunos parámetros de las EQs.

%


\chapter{Estrellas de Quarks magnetizadas.}\label{cap4}
\markright{Capítulo 4: Estrellas de Quarks magnetizadas.}

\section{Observables.}

\subsection{Relación Masa Radio y ecuaciones TOV.}

En el  capítulo anterior estudiamos la estabilidad y las EdE de un
gas magnetizado de quarks. ¿Cómo podemos vincular esto con la
curvatura en el espacio tiempo, causada por las enormes densidades
de la materia en una EQs?

El problema se puede tratar separando la fuerza gravitacional que
tiene un rango de acción infinito y las fuerzas fuertes de corto
alcance. De esta manera en nuestro estudio anterior hemos
considerado que en las escalas de las interacciones fuertes $(\sim
1\,\mathrm{fm})$ la curvatura es nula, la gravitación no es
necesaria incluirla.

Ahora bien, para las escalas de las EQs, $(\sim 10\,\mathrm{km})$ no
podemos obviar el papel de la curvatura, es decir, la gravedad curva
el espacio tiempo a escalas macroscópicas~\cite{weber}.

Por esto para conocer la estructura (radio, masa) de una estrella es
necesario utilizar la Teoría General de la Relatividad de Einstein.

Las ecuaciones de Einstein establecen un estrecho vínculo entre el
contenido de materia en un lugar del espacio y la curvatura del
mismo~\cite{MTW}:

\begin{equation}
G^{\mu}_{\,\,\,\,\nu}=\kappa \mathcal{T}^{\mu}_{\,\,\,\,\nu},
\label{EE1}
\end{equation}
$(\mu,\nu=0,1,2,3)$, $\kappa=8\pi
\mathrm{G_N}$$\mathrm{G_N}=1.32\times
10^{-42}\,\mathrm{fm}\mathrm{MeV}^{-1}$, $\mathrm{G}_{{\mu
\nu}}=R_{{\mu \nu}}-\frac{1}{2}R g_{{\mu \nu}}$ es el tensor de
Einstein el cual viene determinado por el tensor de Ricci $R_{{\mu
\nu}}$ y por el escalar de Ricci $R=R^{\mu}_{\,\,\,\,\mu}$, estos
últimos dependen de segundas derivadas de la métrica.

\begin{equation}
R_{\mu\nu}=\Gamma^{\alpha}_{\mu\nu,\alpha}-\Gamma^{\alpha}_{\mu\alpha,\nu}+\Gamma^{\alpha}_{\mu\nu}
\Gamma^{\beta}_{\alpha\beta}-\Gamma^{\beta}_{\mu\alpha}
\Gamma^{\alpha}_{\nu\beta}, \label{TF2}
\end{equation}

\noindent las cantidades $\Gamma^{\alpha}_{\mu\nu}$ son los índices
de Christoffel, que dependen de primeras derivadas de la métrica por
la fórmula,
\begin{equation}
\Gamma^{\alpha}_{\mu\nu}=\frac{g^{\alpha\beta}}{2}(g_{\beta\mu,\nu}+g_{\nu\beta
,\mu}-g_{\mu\nu,\beta}). \label{TF3}
\end{equation}
El tensor energía impulso $\mathcal{T}^{\mu}_{\,\,\,\,\nu}$ viene
dado por la expresión~(\ref{Tensor_EM_mag}).

Para una estrella estática en equilibrio se emplea la métrica:
\begin{equation}\label{MetricaSch}
    ds^{2} = -e^{2\Phi(r)}dt^{2} + e^{\Lambda(r)}dr^{2} + r^2 d\theta^{2}
    +r^2 \sin^2 \theta d\phi^2
\end{equation}

Utilizando la métrica~(\ref{MetricaSch}), la ecuación~(\ref{EE1}) y
la ley de conservación de la energía
$(\mathcal{T}^{\mu\nu}_{\,\,\,\,;\nu}=0)$ podemos encontrar las
ecuaciones TOV~\cite{MTW}, que nos dan las configuraciones de
estrellas estáticas y con simetría esférica:

\begin{subequations}\label{TOV}
\begin{eqnarray}
  \frac{dM}{dr} &=& 4\pi G\epsilon(r) \\
  \frac{dP}{dr} &=& -G\frac{(\epsilon(r)+P(r))(M(r)+4\pi P(r)r^3)}{r^2-2rM(r)}
\end{eqnarray}
\end{subequations}

Para resolver el sistema de ecuaciones~(\ref{TOV}) utilizaremos las
soluciones numéricas de las EdE para la MEQM~(\ref{TQf})
y~(\ref{Pper}). El radio $R$ y la masa correspondiente $M$ de la
estrella se determinan imponiendo la condición de presión cero
$P(R)=0$. La presión central queda fijada por la EdE, $P(0)=P_c$, si
imponemos además la condición $M(0)=0$ podemos resolver el sistema
de ecuaciones TOV.

\begin{figure}[t!h]
\centering
\begin{minipage}{8cm}
\includegraphics[height=7cm,width=8cm]{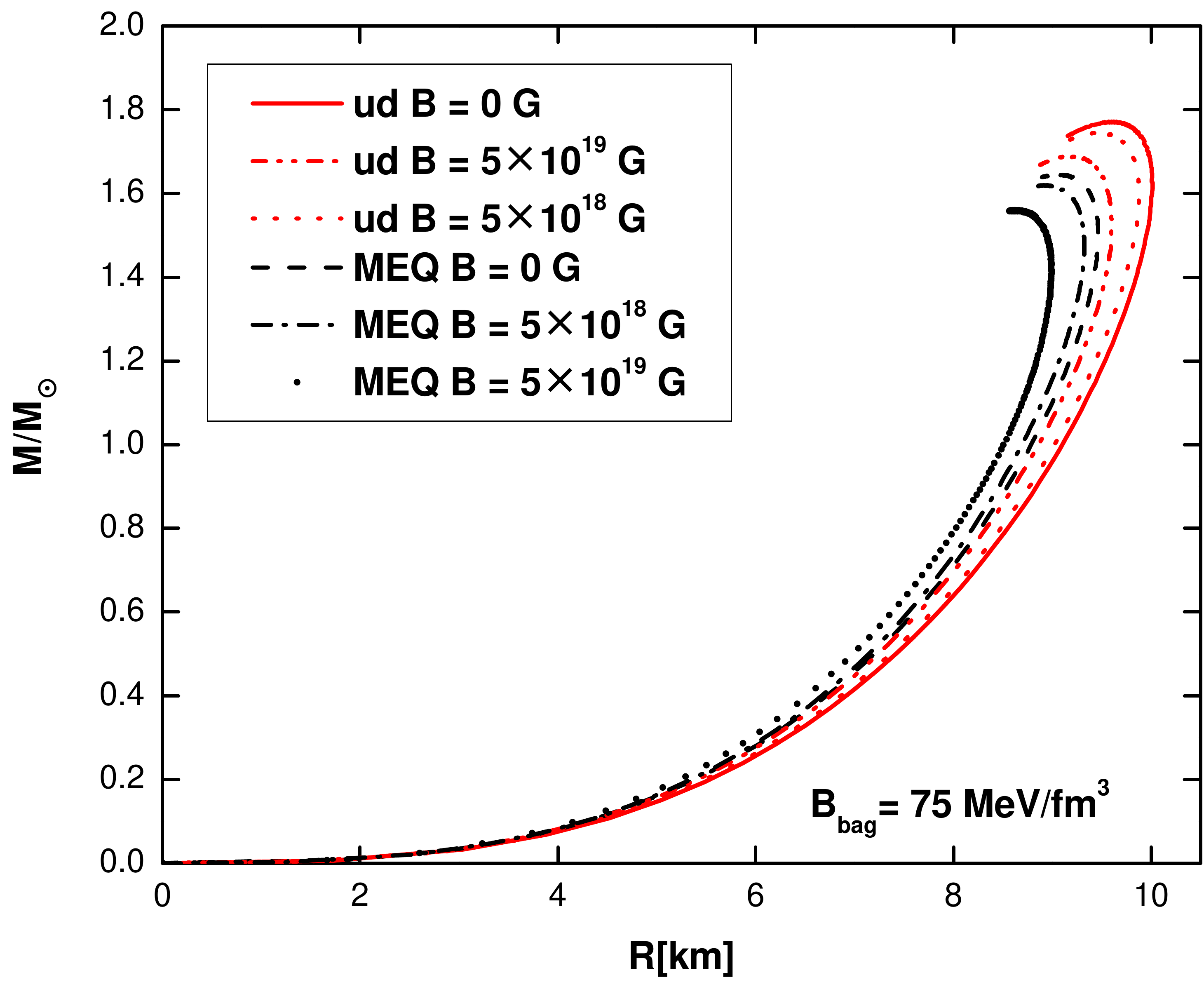}
      \caption{\small Configuraciones de $M-R$ obtenidas para MEQM. Se
      muestran curvas para diferentes valores de campo magnético $B$.
      Para comparar se han añadido las gráficas $M-R$ para la
      materia de quarks normal. Las configuraciones más
      compactas son las de MEQM}
\label{tov}
\end{minipage}
\hfill
\begin{minipage}{8cm}
 \includegraphics[height=7cm,width=8cm]{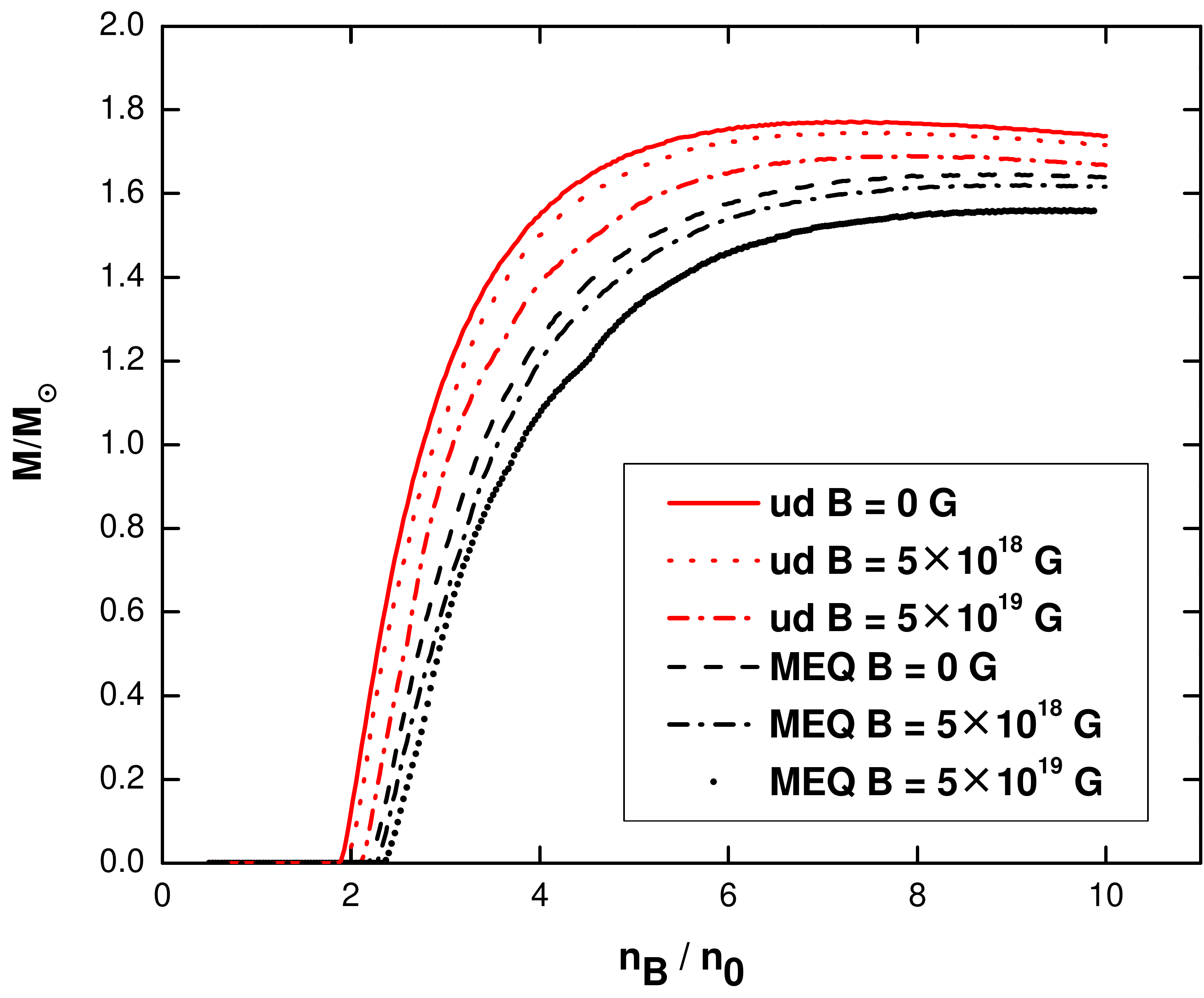}
      \caption{\small  Masa contra densidad central para diferentes valores del campo magnético.
      Se muestran los resultados para el gas formado por $u$ y $d$. Se puede observar que el aumento del campo tiende a
      hacer decrecer la masa máxima para una densidad fija.} \label{rhocvsM}
\end{minipage}
\end{figure}

En la \fig{tov} mostramos los gráficos de las configuraciones de
equilibrio para la MEQ y para la MEQM con un valor del parámetro de
Bag $(B_{bag} = 75\,\mathrm{MeV\,fm}^{-3})$. Como ya fue mencionado
en el capítulo anterior, el campo magnético hace que la MEQ sea más
estable por lo que las estrellas pueden ser más compactas. Se
observa como a medida que aumenta el campo las configuraciones
obtenidas presentan menores radios.

La solución de las ecuaciones TOV~(\ref{TOV}) representadas en la
\fig{tov} dan las configuraciones de equilibrio para la estrella,
pero este equilibrio puede ser estable o inestable.

Para conocer si una configuración es estable o inestable hay que
estudiar la dependencia de la masa con densidad central de masa
$\rho_c$ (equivalentemente $n_B$) \fig{rhocvsM}. La estabilidad de
la estrella puede determinarse por la pendiente de esta curva.
Cuando $dM/d\rho_c<0$ las configuraciones de equilibrio se
consideran inestables~\cite{Weinberg}.

En la \tab{resMR} se muestran los valores de masa máxima con su
correspondiente radio y densidad central para diferentes valores de
campo obtenidos de la \fig{tov} y \fig{rhocvsM}.

\begin{table}[h!]
\begin{center}
\begin{tabular}{|l|c|c|c|c|}
\hline

Materia & $M_{max}$ $(M/M_{\odot})$ & R $(km)$ & B $(G)$ & $n_{Bc}/n_0$ \\
\hline\hline

MEQM & 1.56 & 8.62  & $1\times10^{19}$  & 9.46  \\ \hline

MEQM & 1.62 & 8.91  & $5\times10^{18}$  & 9.14 \\ \hline

MEQ  & 1.65 & 9.07  & 0                 & 8.48 \\ \hline

ud   & 1.69 & 9.20  & $1\times10^{19}$  & 7.88 \\ \hline

ud   & 1.74 & 9.50  & $5\times10^{18}$  & 7.31 \\ \hline

ud   & 1.77 & 9.60  & 0                 & 7.02 \\

\hline
\end{tabular}
\end{center}
\caption{\small Resultados obtenidos de la \fig{tov}.} \label{resMR}
\end{table}

\subsection{Masa Bariónica.}

La masa que encontramos en el epígrafe anterior al resolver las
ecuaciones TOV~(\ref{TOV}) es la masa gravitacional ($M_G$).
Hallemos ahora la masa bariónica~\cite{Bombaci}.  Ella difiere de la
gravitacional porque vamos a considerar que la densidad bariónica
depende de la métrica.

Para hallarla tengamos en cuenta que el volumen de una capa esférica
de la estrella en la métrica de Schwarzschild está dada por la
expresión:

\begin{equation}\label{Vol}
    dV=\frac{4\pi r^2\,dr}{[1-\frac{2G_Nm}{r^2}]^{1/2}},
\end{equation}

el número de bariones dentro de la estrella lo podemos calcular
como:
\begin{equation}\label{Bar}
    N_B=\int n(r)dV=\int_0^R\frac{4\pi r^2\,dr}{[1-\frac{2G_Nm}{r^2}]^{1/2}}
\end{equation}
donde $n(r)$ es el número de bariones por unidad de volumen. Por
tanto la masa bariónica de la estrella es:

\begin{equation}\label{masaB}
    M_B=m_n\cdot N_B. \ \ \ \ 
\end{equation}
Se toma $m_n$ como la masa bariónica $m_n=939\,\mathrm{MeV}$.
\begin{figure}[t!h]
\centering
\begin{minipage}{8cm}
\includegraphics[height=7cm,width=8cm]{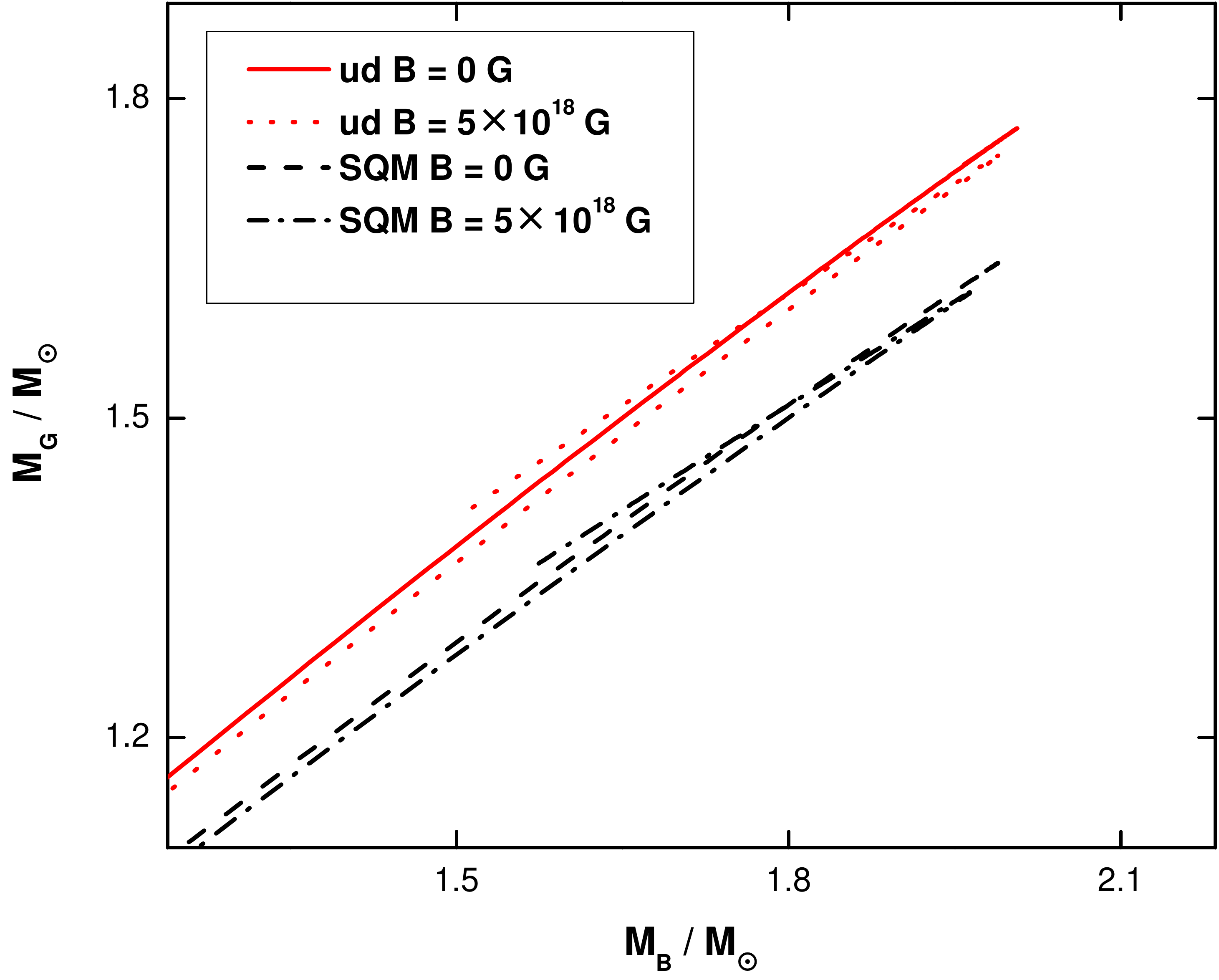}
      \caption{\small Masa bariónica contra masa gravitacional.}\label{MBvsMg}
\end{minipage}
\hfill
\begin{minipage}{8cm}
 \includegraphics[height=7cm,width=8cm]{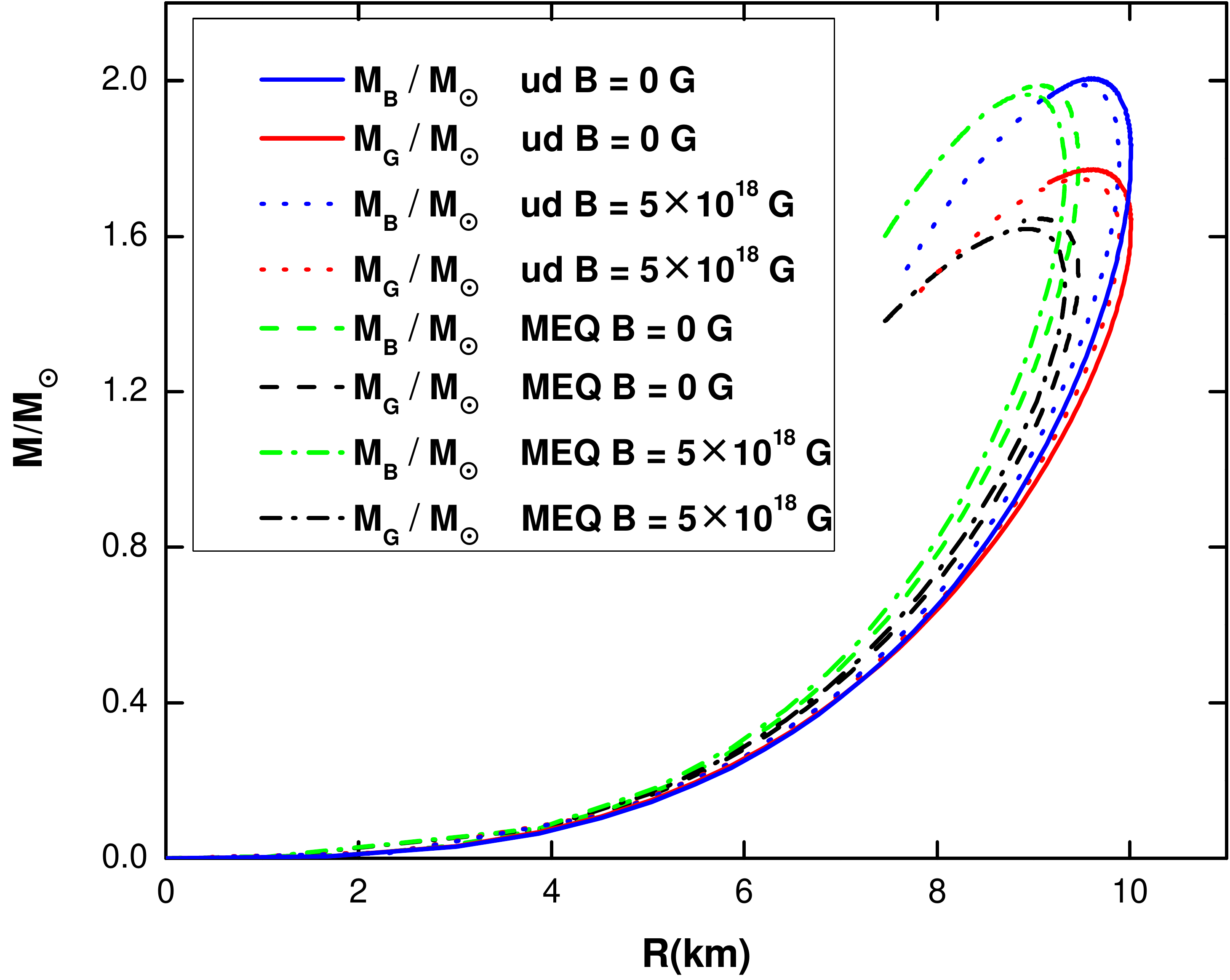}
      \caption{\small Masa bariónica contra radio.}\label{MBvsR}
\end{minipage}
\end{figure}

La \fig{MBvsMg} muestra la masa gravitatoria en función de la masa
bariónica. La \fig{MBvsR} representan las configuraciones de
equilibrio de masa bariónica y radios. Como se puede observar la
masa bariónica es siempre superior a la masa gravitacional. La
estabilidad puede ser estudiada de la misma manera en que se analiza
la estabilidad para la masa gravitacional \cite{Weinberg}. Podemos
apoyarnos en la \fig{MBvsMg} que como se puede observar alcanza un
máximo y luego comienza a decrecer hacia menores valores de masa
bariónica, esta es la zona de inestabilidad.


\subsection{Corrimiento al rojo gravitacional.}


El corrimiento al rojo se define como:

\begin{equation}
1 + z =\frac{f_{(\text{receptor})}}{f_{(\text{fuente})}} \,\,
\Rightarrow \,\, 1 + z
=\sqrt{\frac{g_{tt}(\text{receptor})}{g_{tt}(\text{fuente})}}
\end{equation}

\noindent donde $g_{tt}$ es la componente temporal de la métrica.

Esta magnitud da la diferencia de frecuencia de la luz emitida por
la superficie de la estrella con respecto a la que se mide en un
sistema en reposo respecto a la fuente. En este caso estamos
interesados en el corrimiento al rojo gravitacional, es decir, el
cambio de frecuencia que sufre la radiación al moverse en un campo
gravitatorio muy intenso como el que existe en las EQs.
\begin{figure}[!ht]
      \centering
      \includegraphics[height=7cm,width=8cm]{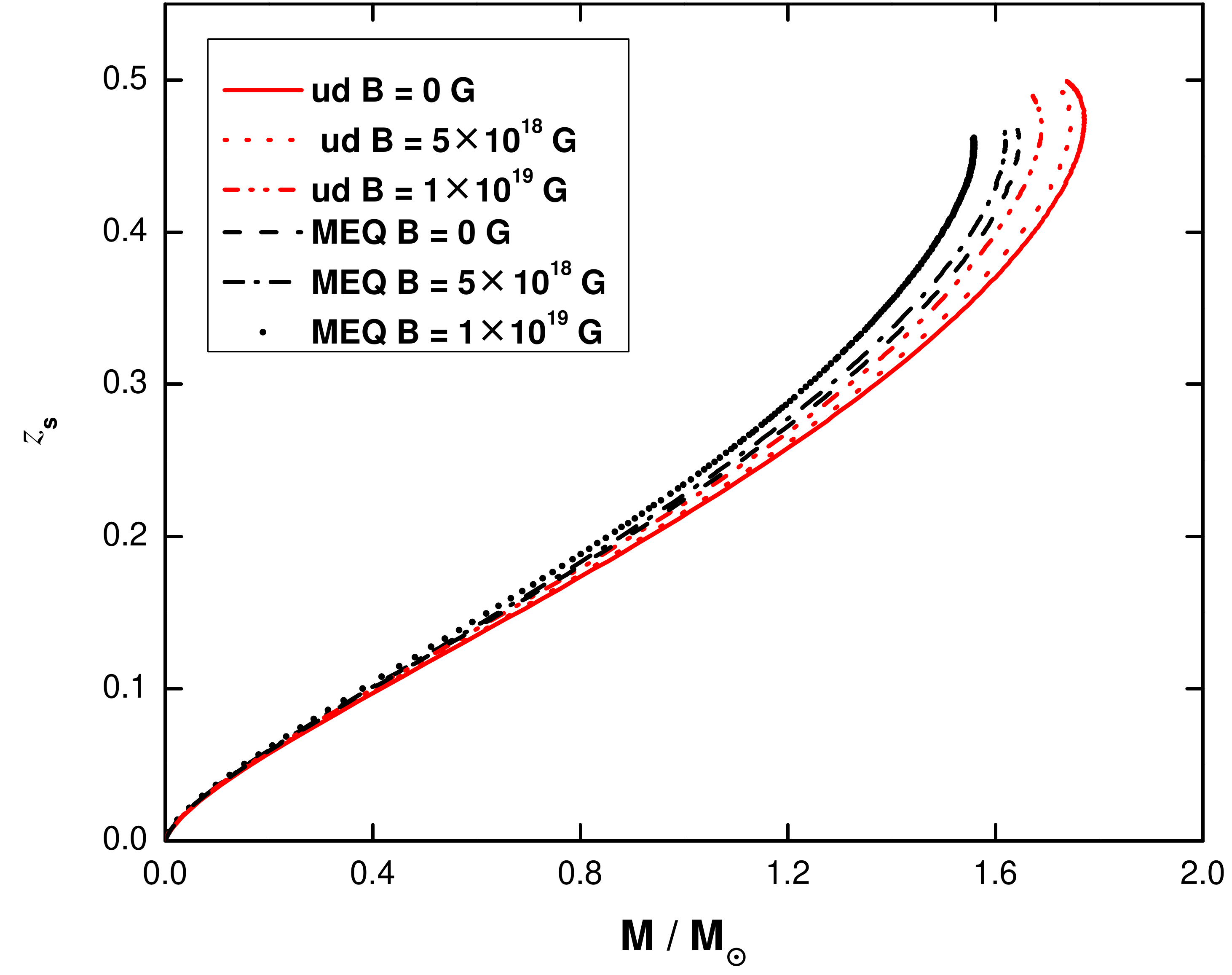}
      \caption{Corrimiento al rojo gravitacional para distintos valores de campo magnético y $B_{bag}=75\,\mathrm{MeV\,fm}^{-3}$.}
\label{redshiftgraf}
\end{figure}

Para el caso de la métrica~(\ref{MetricaSch}) se puede obtener
que~\cite{Weinberg}:
\begin{equation}\label{redshift}
    z_s=\frac{1}{\sqrt{1-2G_NM/R}}-1
\end{equation}

El corrimiento al rojo se utiliza para descartar EdE que no
reproduzcan los datos observados. En la \fig{redshiftgraf} mostramos
los resultados del corrimiento al rojo gravitacional. El valor de
$z_s\simeq0.345$ obtenido experimentalmente en las líneas
espectrales de rayos X proveniente del objeto EXO 0748-676 se
corresponderá en nuestro caso con valores de $M/M_{\odot}\simeq1.42$
resultado que está dentro de los limites de masas medidas
experimentalmente.

%


\chapter*{Conclusiones.}\label{cap5}
\addcontentsline{toc}{chapter}{Conclusiones.}

\markright{Conclusiones.}

En este tesis se ``tiende el puente'' entre el estudio microscópico
de la  materia de quarks  magnetizada (dado por el estudio de las
EdE y la estabilidad) y los  observables macroscópicos de las EQs
derivados de las EdE de ese tipo de materia magnetizada.

El estudio se hace tanto para la MEQ, como para la normal en
equilibrio estelar en presencia de un campo magnético intenso,
utilizando el modelo fenomenológico de Bag del MIT.

La estabilidad se ha estudiado con el propósito de conocer si el
campo magnético contribuye o no  a que sea  más estable la materia
de quarks. Las EdE se han obtenido para utilizarlas al resolver las
ecuaciones de equilibrio hidrodinámico TOV y obtener cómo los
observables (masa gravitacional, radios y  corrimiento al rojo), se
modifican con el campo magnético.

Los resultados obtenidos pueden resumirse de la siguiente manera:

\begin{enumerate}
\item Se obtuvo numéricamente la EdE para la MEQM \fig{EoS} y para la materia
normal de quarks en equilibrio estelar.

\item Se estudió la estabilidad de la MEQM en equilibrio estelar:
\begin{itemize}
  \item Se verificó que las desigualdades de estabilidad~(\ref{stabineq})
se cumplen para un amplio rango de parámetros de modelo de Bag.
\item La MEQM es más estable que la MEQ y que la materia de quarks
normal.
  \item Se obtuvieron las ventanas de estabilidad: curvas en el plano
$n/n_B$ contra $m_s$ de la MEQM teniendo en cuenta la variación de
la masa del quark $s$, la densidad bariónica, el campo magnético y
el parámetro de Bag.
  \item Mientras que para la MEQ el rango de la densidad bariónica permitido
está entre $1.8 \lesssim n_B/n_0 \lesssim 2.4$ para masas del quark
s en el rango $50 \leq m_s \leq 300\,\mathrm{MeV}$, la MEQM admite
densidades en el rango $1.85 \lesssim n_B/n_0 \lesssim 2.6$ para
rangos de la $m_s$ $50 \leq m_s \leq 300\,\mathrm{MeV}$ y campo
magnético de $5\times 10^{18}$~G. El rango del parámetro de Bag es
$57\lesssim B_{bag} \lesssim 90\,\mathrm{MeV\,fm^{-3}}$.
\end{itemize}
\item Se resolvieron las ecuaciones TOV con las EdE obtenidas.
La relación masa-radio (MR) muestra como el  campo magnético
contribuye a soluciones estables de EQs más compactas con  menores
valores de masa gravitacional y radio.
\begin{itemize}
  \item Se obtuvo los efectos del campo magnético en la masa bariónica.
 Se resolvieron las ecuaciones TOV
para ella y se obtuvieron las relaciones masa bariónica en función
del radio.
  \item Se estudió el corrimiento al rojo de las EQs magnetizadas.
\end{itemize}
\end{enumerate}

\newpage

\chapter*{Recomendaciones.}
\addcontentsline{toc}{chapter}{Recomendaciones.}
\markright{Recomendaciones.}

Estudios teóricos recientes han mostrado que a densidades muy altas
la materia de quarks podría estar en una fase superconductora de
color: \emph{Color Flavor Locked }(CFL) siendo éste el estado
fundamental de la materia \cite{Rajagopal:2000ff}. Estudiar como
influye la presencia del campo magnético en la estabilidad de esta
fase resulta un tema relevante que ya hemos emprendido.

Por otro lado los observables astrofísicos aquí discutidos merecen
que sean comparados con datos observacionales.  Es esta otra
dirección de trabajo en que pensamos dirigir nuestros esfuerzos.
Estudiar otros observables como el momento de inercia de la estrella
es también algo que ha quedado pendiente.

La inclusión de los efectos de la rotación de las EQs es un tema
digno de analizarse,  los efectos del campo magnético pueden en este
caso resultar importantes.

%



\begin{thebibliography}{99}
\addcontentsline{toc}{chapter}{Bibliografía.}
\markright{Bibliografía.}

\bibitem{MTW} Charles W. Misner, KipS. Thorne, John. Archibald Wheeler, {\it Gravitation},
ed W.H.Freeman and Company, NY (1973). 

\bibitem{Baym:2006rq}
  G.~Baym,
  AIP Conf.\ Proc.\  {\bf 892}, 8 (2007)
  [arXiv:nucl-th/0612021].

\bibitem{Shapiro} S.~L.~Shapiro \& S.~A.~Teukolsky, 1983, Black
Holes , White Dwarfs and Neutron Stars (New York: John Wiley \&
Sons).

\bibitem{Lattimer:2004pg}
  J.~M.~Lattimer and M.~Prakash,
  Science {\bf 304}, 536 (2004)
  [arXiv:astro-ph/0405262].

\bibitem{Bombaci:2008kb}
  I.~Bombaci,
  arXiv:0809.4228 [gr-qc].

\bibitem{SchaffnerBielich:2004ch}
  J.~Schaffner-Bielich,
  J.\ Phys.\ G {\bf 31}, S651 (2005)
  [arXiv:astro-ph/0412215].



\bibitem{Roy:1999rc}
  D.~P.~Roy,
  arXiv:hep-ph/9912523.


\bibitem{PDG}
Journal of Physics G: Nuclear and Particle Physics Vol 33 July 2006
Pp 1-1232 (Complete volume) Review of particle Physics

\bibitem{Bodmer:1971we}
  A.~R.~Bodmer,
  Phys.\ Rev.\  D {\bf 4}, 1601 (1971).

\bibitem{Witten:1984rs}
  E.~Witten,
  ``Cosmic Separation Of Phases,''
  Phys.\ Rev.\  D {\bf 30}, 272 (1984).

\bibitem{Lattimer:2006xb}
  J.~M.~Lattimer and M.~Prakash,
  Phys.\ Rept.\  {\bf 442}, 109 (2007)
  [arXiv:astro-ph/0612440].

\bibitem{weber} For a recent review on strange quark matter and compact stars see
: F. Weber, Prog. Part. Nucl. Phys. {\bf 54}, 193 (2005).

\bibitem{Felipe:2007vb}
  R.~Gonz\'{a}lez~Felipe, A.~P\'{e}rez~Mart\'{\i}nez, H.~P\'{e}rez~Rojas and M.~G.~Orsaria,
  Phys.\ Rev.\  C {\bf 77}, 015807 (2008).

\bibitem{Felipe:2008cm}
  R.~Gonz\'{a}lez~Felipe, A.~P\'{e}rez~Mart\'{\i}nez,
  J.\ Phys.\ G {\bf 36}, 075202 (2009).

\bibitem{PerezMartinez:2005av}
  A.~Perez Martinez, H.~Perez Rojas, H.~J.~Mosquera Cuesta, M.~Boligan and M.~G.~Orsaria,
  Int.\ J.\ Mod.\ Phys.\  D {\bf 14}, 1959 (2005)
  [arXiv:astro-ph/0506256].


\bibitem{Oppenheimer} J.~R.~Oppenheimer and G.~M.~Volkoff,  Phys. Rev. {\bf 55}, 374 (1939).

\bibitem{Ruester:2006yh}
  S.~B.~Ruester,
  arXiv:nucl-th/0612090.

\bibitem{Duncan1} R.~C.~Duncan \& C.~Thompson, ApJL {\bf 392}, L9 (1992),

\bibitem{Duncan2} R.~C.~Duncan \& C.~Thompson, ApJ {\bf 469}, 764 (1996)

\bibitem{Lattimer}
  J.~M.~Lattimer and M.~Prakash.
  \emph{Astrophys}.~J.~550 (2001) 426.  {\bf 304}, 536 (2004)

\bibitem{12} A. Perez Martinez, H. Perez Rojas, H. Mosquera Cuesta. \emph{Chin.
Phys. Lett}. Vol 21 No. {\bf 11} (2004) 2117-2119.

\bibitem{itoh}
  N.~Itoh,
   Prog.\ Theor.\ Phys.\  {\bf 44}, 291 (1970).


\bibitem{Usov}V. V. Usov, Phys. Rev. Lett. {\bf 87}, 021101 (2001).

\bibitem{Hewish}
A.~Hewish, S.~J.~Bell, J.~D.~H.~Pilkington, P.~F.~Scott \&
R.~A.~Collins,
Nature {\bf217}, 709 - 713 (1968)


\bibitem{Gold}
T.~Gold,
Nature {\bf221}, 25 - 27 (1969)

\bibitem{Drake:2002bj}
  J.~J.~Drake {\it et al.},
  Astrophys.\ J.\  {\bf 572}, 996 (2002)
  [arXiv:astro-ph/0204159].


\bibitem{Weisberg:2004hi}
  J.~M.~Weisberg and J.~H.~Taylor,
  ASP Conf.\ Ser.\  {\bf 328}, 25 (2005)
  [arXiv:astro-ph/0407149].

\bibitem{Xu:2001bp}
  R.~X.~Xu and G.~J.~Qiao,
  arXiv:astro-ph/0108235.

\bibitem{Xu:2002ns}
  R.~X.~Xu,
  Chin.\ J.\ Astron.\ Astrophys.\  {\bf 3}, 33 (2003)
  [arXiv:astro-ph/0211214].

\bibitem{Trumper:2003we}
  J.~E.~Trumper, V.~Burwitz, F.~Haberl and V.~E.~Zavlin,
  Nucl.\ Phys.\ Proc.\ Suppl.\  {\bf 132}, 560 (2004)
  [arXiv:astro-ph/0312600].


\bibitem{Ozel:2006km}
  F.~Ozel,
  arXiv:astro-ph/0605106.



\bibitem{Xu:2007wd}
  R.~Xu,
  AIP Conf.\ Proc.\  {\bf 968}, 197 (2008)
  [arXiv:0709.1305 [astro-ph]].


\bibitem{Buballa}
M.~Buballa, NJL-model analysis of dense quark matter,
[arXiv:hep-ph/0402234v2] 27 Jan. 2005.

\bibitem{Xu:2008nd}
  R.~Xu,
  J.\ Phys.\ G {\bf 36}, 064010 (2009)
  [arXiv:0812.4491 [astro-ph]].

\bibitem{Chodos}
A.~Chodos, R.~L.~Jaffe, K.~Johnson, C.~B.~Thorn, and
V.~F.~Weisskopf, Phys. Rev. D 9 (1974) 3471.

\bibitem{Madsen:1999ci}
  J.~Madsen,
  Phys.\ Rev.\ Lett.\  {\bf 85}, 10 (2000).
  [arXiv:astro-ph/9912418].

\bibitem{Schmitt:2010pn}
  A.~Schmitt,
  arXiv:1001.3294 [astro-ph.SR].



\bibitem{Chakrabarty:1996te} S.~Chakrabarty, Phys.\ Rev.\  D {\bf 54}, 1306 (1996).

\bibitem{Landau PHE} Course of theoretical Physics. Statistical Physics p.I y p.II L. D. Landau
\& E. M. Lifshitz. Pergamon International Library.

\bibitem{CAMAYTE} Carlos Rodríguez Castellanos, María Teresa Pérez Maldonado, Introducción a la Física Estadística, ed Félix Varela, La Habana (2002).

\bibitem{Bagrov} V.~G.~Bagrov, D.~M.~Gitman, {\it Exact solutions of
relativistic wave equations} (Kluwer Academic Publ (1990).

\bibitem{Weinberg} Steven Weinberg, 1972, Gravitation and Cosmology: Principles and Aplications of the
General Theory of Relativity (New York: John Wiley \& Sons).

\bibitem{Bombaci} I.~Bombaci,
Astron.\&\ Astrophys.  {\bf305}, 871-877 (1996)


\bibitem{Rajagopal:2000ff}
  K.~Rajagopal and F.~Wilczek,
  Phys.\ Rev.\ Lett.\  {\bf 86}, 3492 (2001).



%




%
%
%
%
%
%
%
%
%
%
%
%
%
%
%
%
%
%
%
%
%
%
%
%
%
%
%
%
%
%
%
%
%
%
%
%
%
%
%
%
%
%
%
%
%
%
%
%
%
%
%
%
%
%
%
%
%
%
%
%
%
%
%
%
%
%
%
%
%
%
%
%
%
%
%
%
%
%
%
%
%
%
%
%
%
%








\end{thebibliography}
\end{document}